\documentclass{aa}
\usepackage{txfonts}
\usepackage{graphicx}

\defcitealias{emsenhuberNewGenerationPlanetary2021}{NGPPS~I}
\defcitealias{emsenhuberNewGenerationPlanetary2021a}{NGPPS~II}

\begin{document}

	\title{The interplay between pebble and planetesimal accretion in population synthesis models and its role in giant planet formation}
	
	\author{A. Kessler
	    \and Y. Alibert
	    }
		
	\institute{Physikalisches Institut, University of Bern, Gesellschaftsstrasse 6, 3012 Bern, Switzerland\\
	    \email{andrin.kessler@unibe.ch}}
	
	\date{Received TBD; Accepted TBD}
	
	\abstract{In the core accretion scenario of planet formation, rocky cores grow by first accreting solids until they are massive enough to accrete gas. For giant planet formation this means that a massive core must form within the lifetime of the gas disk. Inspired by observations of solar system features such as the asteroid and Kuiper belts, the accretion of roughly km-sized planetesimals is traditionally considered as the main accretion mechanism of solids but such models often result in longer planet formation timescales. The accretion of mm-cm sized pebbles, on the other hand, allows for rapid core growth within the disk lifetime. The two accretion mechanisms are typically discussed separately.}
	{We investigate the interplay between the two accretion processes in a disk containing both pebbles and planetesimals for planet formation in general and in the context of giant planet formation specifically. The goal is to disentangle and understand the fundamental interactions that arise in such \emph{hybrid} pebble-planetesimal models laying the groundwork for informed analysis of future, more complex, simulations.}
	{We combine a simple model of pebble formation and accretion with a global model of planet formation which considers the accretion of planetesimals. We compare synthetic populations of planets formed in disks composed of different amounts of pebbles and 600 meter sized planetesimals to identify the impact of the combined accretion scenario. On a system-level, we study the formation pathway of giant planets in these disks.}
	{We find that, in hybrid disks containing both pebbles and planetesimals, the formation of giant planets is strongly suppressed whereas in a pebbles-only or planetesimals-only scenario, giant planets can form. We identify the heating associated with the accretion of up to 100 km sized planetesimals after the pebble accretion period to delay the runaway gas accretion of massive cores. Coupled with strong inward type-I migration acting on these planets, this results in close-in icy sub-Neptunes originating from the outer disk.}
	{We conclude that, in hybrid pebble-planetesimal scenarios, the late accretion of planetesimals is a critical factor in the giant planet formation process and that inward migration is more efficient for planets in increasingly pebble dominated disks. We expect a reduced occurrence rate of giant planets in planet formation models that take into account the accretion of pebbles and planetesimals.}

	
	\titlerunning{The interplay between pebble and planetesimal accretion}
	\authorrunning{A. Kessler et al.}
	\maketitle
	
	\section{Introduction}
	    
	    In the core accretion paradigm, the formation of giant planets is inherently constrained by the lifetime of the circumstellar gas disk. A protoplanet core must grow massive enough on the time scale of a few Myr in order to accrete significant amounts of gas before the dispersal of the disk \citep{haischDiskFrequenciesLifetimes2001}. Classical planet formation models consider the accretion of planetesimals \citep[e.g.][]{pollackFormationGiantPlanets1996,alibertModelsGiantPlanet2005}. From the size frequency distribution of solar system asteroids, the diameter of primordial planetesimals is estimated to be around 100 km \citep{bottkeFossilizedSizeDistribution2005, morbidelliAsteroidsWereBorn2009}. This is supported by simulations of planetesimal formation through the streaming instability \citep{schaferInitialMassFunction2017}. On the other hand, observations of small Kuiper belt objects suggest a larger number of kilometer sized planetesimals \citep{arimatsuKilometresizedKuiperBelt2019} which is consistent with small primordial planetesimals \citep{schlichtingINITIALPLANETESIMALSIZES2013}. The exact size distribution of primordial planetesimals remains uncertain. Core growth time scales using large planetesimals are long, typically exceeding the disk lifetime \citep{pollackFormationGiantPlanets1996}. However, giant planet formation is shown to be successful in a planetesimals-only setting when sub-kilometer sized planetesimals are considered \citep[][hereafter \citetalias{emsenhuberNewGenerationPlanetary2021}]{emsenhuberNewGenerationPlanetary2021}. Even smaller, roughly mm-cm sized, objects called pebbles are more strongly affected by gas drag and can be captured efficiently, forming cores quickly  \citep{ormelEffectGasDrag2010, lambrechtsRapidGrowthGasgiant2012}. Planet formation models considering the accretion of such pebbles typically produce giant planets comparatively easily while disregarding planetesimal-like objects entirely \citep{lambrechtsFormingCoresGiant2014,bitschGrowthPlanetsPebble2015,bruggerMetallicityEffectPlanet2018}.\\
	    
	    High-precision measurements of isotopes in meteorites suggest that the early population of small bodies in the Solar System has been separated into two reservoirs for $\sim$2-3 Myr. The origin of this dichotomy is unknown. The forming Jupiter is theorised to have acted as a radial barrier for planetesimals \citep{kruijerAgeJupiterInferred2017a,brasserPartitioningInnerOuter2020} whereas other proposed explanations link the dichotomy to protoplanetary disk effects such as pressure bumps related to silicate and volatile evaporation fronts \citep{lichtenbergBifurcationPlanetaryBuilding2021,izidoroPlanetesimalRingsCause2021,morbidelliContemporaryFormationEarly2021}. In order for proto-Jupiter to separate the drifting pebbles for several millions of years, the core mass must remain at least around 20 Earth masses. In any standard planet formation model this is very unlikely to happen. In this mass range, the gravitational pull of the planet triggers rapid gas accretion, quickly forming a Jupiter-like planet. In \citet{alibertFormationJupiterHybrid2018}, a Jupiter formation scenario using a combination of pebble and planetesimal accretion is suggested to connect the Jovian formation history to the observational constraints. They consider the \emph{in-situ} formation of Jupiter in their proof of concept study. Fast core growth to $\sim$10-20 Earth masses within $\sim$1 Myr is facilitated by the accretion of pebbles. At this mass, the further accretion of pebbles is prevented by a pressure bump outside the planetary orbit \citep{paardekooperDustFlowGas2006,lambrechtsFormingCoresGiant2014}. Slow planetesimal accretion can sufficiently heat the envelope for the pressure to balance the gravitational pull on the surrounding gas, delaying runaway gas accretion. At some point, Jupiter grows massive enough to quickly accrete large amounts of gas, reaching its present day mass. \\
	    
	    Motivated by this proposed formation scenario of Jupiter, we investigate the consequences of a combined pebble and planetesimal accretion model for the formation of giant planets and planet formation in general. We modify the Bern model of planet formation and evolution \citepalias{emsenhuberNewGenerationPlanetary2021} with a simple model of pebble formation and accretion \citep{bitschGrowthPlanetsPebble2015,bruggerMetallicityEffectPlanet2018}. The Bern model of planet formation and evolution is a global model that self-consistently computes the evolution of the gas disk, the dynamics of the planetesimal disk, the accretion of gas and planetesimals by planetary embryos, the planet-planet N-body interactions, as well as planet-gas interactions such as gas-driven migration.\\
	    
	    Population synthesis allows to probe a large part of the parameter space of planet formation. For this reason, we investigate the effects of the two solid accretion mechanisms on a population level. The primary goal here is to understand the interplay of classical planetesimal accretion and pebble accretion models, not to predict the characteristics of a physical population of planets formed from a disk consisting of pebbles and planetesimals. To remove the chaotic component of multi-planetary systems, we investigate the formation of a single planet per disk. This allows us to isolate the key differences in the formation pathway of planets forming in disks composed of pebbles \emph{and} planetesimals. A comparison to the observed population of planetary systems would necessarily require the simultaneous modelling of multiple planets per disk. To uncover the interplay of the two accretion mechanisms, we vary the amount of pebbles with respect to planetesimals. We especially focus on giant planet formation as a key topic of interest in the scientific debate about the size of the accreted solids.\\
	    
	    In Sect. \ref{sec:theoretical models}, we give a brief overview of our planet formation and evolution model. The pebble formation and accretion model is described in more detail. In Sect. \ref{sec:population synthesis outcomes}, we present the populations emerging from different solid disk compositions. We compare populations from a pure planetesimal disk, a pebble-poor (30\%), a pebble-rich (70\%), and a pure pebble disk. We focus on the formation of giant planets in Sect. \ref{sec:giant planet formation}. Particularly, we investigate the onset of rapid gas accretion as well as the impact of orbital migration and the pebble isolation mass. Finally, Sect. \ref{sec:discussion and conclusion} is dedicated to a brief summary of the results and conclusions.
		
	\section{Theoretical models} \label{sec:theoretical models}
	
		We first give a short overview of the model components outlined in \citet{bruggerPebblesPlanetesimalsOutcomes2020} and described in great detail in \citetalias{emsenhuberNewGenerationPlanetary2021}. In particular, we detail the gas disk model, the treatment of planetesimals, the gas accretion model, and the planetary migration prescriptions. We then present the pebble formation model, and finally, the pebble accretion model in more detail.
	
		\subsection{Gas disk model} \label{sec:gas disk model}
			
		  The time evolution of the protoplanetary gas disk surface density $\Sigma_{\text{gas}}$ is governed by the 1D radially symmetric viscous diffusion equation \citep{lustEntwicklungUmZentralkorper1952,lynden-bellEvolutionViscousDiscs1974}
                \begin{equation}
			    \frac{\partial \Sigma_{\text{gas}}}{\partial t} = \frac{1}{r}\frac{\partial}{\partial r}\left[ 3r^{1/2}\frac{\partial}{\partial r}\left( r^{1/2}\nu \Sigma_{\text{gas}}\right)\right] - \dot{\Sigma}_{\text{gas,ph}} - \dot{\Sigma}_{\text{gas,pl}}, \label{eq:gas disk evo}
			\end{equation}
            where $r$ is the orbital distance, $\dot{\Sigma}_{\text{gas,ph}}$ is the sink term related to internal and external photo-evaporation following \citet{mordasiniCharacterizationExoplanetsTheir2012}, and $\dot{\Sigma}_{\text{gas,pl}}$ is the sink term due to the accretion of gas by planets. We use the $\nu = \alpha c_sH$ viscosity parametrisation of \citet{shakuraBlackHolesBinary1973}, where $c_s$ is the isothermal sound speed and $H=c_s/\Omega_K$ is the vertical scale height at Kepler frequency $\Omega_K=\sqrt{GM_\star/r^3}$, $G$ being the gravitational constant and $M_\star$ the stellar mass. In this work we set $\alpha=0.002$ \citep[][hereafter \citetalias{emsenhuberNewGenerationPlanetary2021a}]{emsenhuberNewGenerationPlanetary2021a}. The initial conditions of the simulations are further described in Sect. \ref{sec:population synthesis outcomes}.\\
			
            The initial radial surface density profile is given by \citep{andrewsPROTOPLANETARYDISKSTRUCTURES2010}
		      \begin{equation}
			    \Sigma_{\text{gas}}(r) = \Sigma_0 \left(\frac{r}{5.2 \; \text{AU}}\right)^{-\beta} \exp\left[-\left(\frac{r}{R_\text{char}}\right)^{(2-\beta)}\right]\left(1-\sqrt{\frac{R_\text{in}}{r}}\right). \label{eq:init gas profile}
			\end{equation}
		  Here, $\Sigma_0$ is the initial gas surface density at 5.2 AU, $\beta$ is fixed to 0.9, $R_\text{char}$ is the characteristic radius, and $R_\text{in}$ is the inner radius where the disk is truncated by the stellar magnetic field \citepalias[see][]{emsenhuberNewGenerationPlanetary2021a}. Typical values of these parameters are $R_\text{in}=0.05$ AU and $R_\text{char}=70$ AU.\\
			
		  The midplane temperature $T_\text{mid}$ is calculated in a semi-analytical approach considering viscous heat dissipation and direct stellar irradiation \citep{nakamotoFormationEarlyEvolution1994,huesoEvolutionProtoplanetaryDisks2005}
			\begin{equation}
			    T_\text{mid}^4 = \frac{1}{2\sigma_\text{SB}}\left( \frac{3}{8} \kappa_R \Sigma_\text{gas} + \frac{1}{2\kappa_P\Sigma_\text{gas}}\right)\dot{E}_\nu + T_\text{irr}^4, \label{eq:t_mid}
			\end{equation}
		  where $\sigma_\text{SB}$ is the Stefan-Boltzmann constant and $\dot{E}_\nu = \frac{9}{4}\Sigma_\text{gas}\nu\Omega_K^2$ is the viscous energy dissipation rate. The Rosseland mean opacity $\kappa_R$ is obtained from the minimum of the grain-free gas opacities of \citet{freedmanGASEOUSMEANOPACITIES2014} and the full interstellar opacities of \citet{bellUsingFUOrionis1994} obtained for a micrometer dust-to-gas ratio of 1\%. For the Planck opacity $\kappa_P$, we follow \citet{nakamotoFormationEarlyEvolution1994}. In reality, the disk opacities are coupled to the evolution of dust which then influences the disk temperature and density evolution. We refer to \citetalias{emsenhuberNewGenerationPlanetary2021} for a more detailed discussion of disk opacity. The temperature due to stellar irradiation $T_\text{irr}$ depends on the stellar temperature $T_\star$, radius $R_\star$, and luminosity $L_\star$ via \citep{adamsDisksTauriStars1988,rudenDynamicalEvolutionProtosolar1991a,chiangSpectralEnergyDistributions1997,huesoEvolutionProtoplanetaryDisks2005}
			\begin{equation}
			    T_\text{irr}^4 = T_\star^4\left[ \frac{2}{3\pi}\left(\frac{R_\star}{r}\right)^3 + \frac{1}{7}\left(\frac{R_\star}{r}\right)^2 \frac{H}{r}\right] + \frac{L_\star}{16\pi r^2 \sigma_\text{SB}}e^{-\tau_\text{mid}} + T_\text{c}^4. \label{eq:T_irr}
			\end{equation}
		  The stellar parameters are obtained from the stellar evolution tracks of \citet{baraffeNewEvolutionaryModels2015}. In this way, the temporal evolution of the star affects the evolution of the disk temperature profile. The stellar luminosity term accounts for the direct irradiation contribution through the midplane, considering the optical depth $\tau_\text{mid}$ through the midplane \citepalias{emsenhuberNewGenerationPlanetary2021}. The term with $T_\text{c}=10\,\text{K}$ adds the heating due to the surrounding molecular cloud.

		\subsection{Planetesimals} \label{sec:planetesimals}
			
			We divide the solids in the disk, given by the initial total solids-to-gas ratio Z\textsubscript{tot}, into planetesimals and dust such that $Z_\text{tot} = Z_\text{plan} + Z_\text{dust}$, as well as a 0.01 M\textsubscript{E} embryo. The planetesimals are described by a surface density with a dynamical state given by their root mean square eccentricity and inclination \citep{fortierPlanetFormationModels2013}. The planetesimal disk evolves considering the effects of aerodynamic drag \citep{adachiGasDragEffect1976,inabaHighAccuracyStatisticalSimulation2001,rafikovFastAccretionSmall2004}, dynamical stirring by protoplanets \citep{guileraConsequencesSimultaneousFormation2010} and by other planetesimals \citep{ohtsukiEvolutionPlanetesimalVelocities2002}. Initially, the planetesimal surface density profile is steeper than the gas disk profile \citep{lenzPlanetesimalPopulationSynthesis2019,drazkowskaPlanetesimalFormationStarts2017} and the planetesimals are in dynamical equilibrium with respect to their self-stirring. We consider rocky and icy planetesimals inside and outside the water ice line respectively. Due to sublimation in the inner parts of the disk, there is a significant decrease in the planetesimal surface density just inside the ice line. Hence the growth via the accretion of planetesimals is most efficient just outside the ice line. The planetesimal disk is initialised such that $Z_\text{plan}$ is the total planetesimals-to-gas ratio. In Fig. \ref{fig:profiles}, we show the initial radial gas and planetesimal surface density profiles of the most and least massive planetesimal disks.\\
			
			The planetesimal accretion rate $\dot{M}_\text{plan}$ of a planetary embryo depends on the Kepler frequency $\Omega_K$, the embryo mass over stellar mass ration $M/M_\star$, the surface density of planetesimals $\Sigma_\text{plan}$, as well as the collision probability of planetesimals $p_\text{coll}$ \citep{chambersSemianalyticModelOligarchic2006}. It is given by
			\begin{equation}
			    \dot{M}_\text{plan} = \Omega_K \bar{\Sigma}_\text{plan} R_H^2 p_\text{coll}, \label{eq:plan acc rate}
			\end{equation}
			where $R_H = r \left(\frac{M}{3 M_\star}\right)^{1/3}$ is the Hill radius and $\bar{\Sigma}_\text{plan}$ is the mean planetesimal surface density in the planet's feeding zone. The feeding zone is centered around the planet with a radius $R_\text{feed}=5R_H$ \citep{fortierPlanetFormationModels2013} for circular orbits. This is always the case in a single planet system.\\
			The collision probability is a function of the planetesimal dynamical state \citep{inabaHighAccuracyStatisticalSimulation2001, chambersSemianalyticModelOligarchic2006} and the capture radius of the protoplanet, which is enhanced by the presence of an envelope \citep{inabaEnhancedCollisionalGrowth2003}. The increased capture radius over the physical radius is crucial for the overall planetesimal accretion rate \citep{podolakInteractionsPlanetesimalsProtoplanetary1988, venturiniJupiterHeavyelementEnrichment2020}. Especially for smaller planetesimals, this means the calculation of gas accretion can not be omitted at any stage of the simulation.
			
		\subsection{Gas accretion model} \label{sec:gas accretion model}
			
		  The gas accretion is calculated by solving the 1D radially symmetric internal structure equations \citep{bodenheimerCalculationsAccretionEvolution1986d} which describe mass conservation, hydrostatic equilibrium, and energy transport respectively
			\begin{eqnarray}
			    \frac{\partial M}{\partial r} & = & 4\pi r^2 \rho, \\ \label{eq:mass conservation}
			    \frac{\partial P}{\partial r} & = & - \frac{GM}{r^2}\rho, \\ \label{eq:hydrostatic eq}
			    \frac{\partial T}{\partial r} & = & \frac{T}{P} \frac{\partial P}{\partial r} \min(\nabla_\text{ad}, \nabla_\text{rad}) , \label{eq:energy transport} 
			\end{eqnarray}
		  where $M$ is the mass enclosed in a sphere of radius $r$, $P$ is the pressure, and $T$ is the temperature. The density $\rho(P,T)$ is obtained from the equations of state of \citet{saumonEquationStateLowMass1995}. In convective zones, the temperature gradient is given by the adiabatic gradient $\nabla_\text{ad}$ from the equations of state. Otherwise we use the radiative gradient \citep{kippenhahnStellarStructureEvolution1990}
			\begin{equation}
			    \nabla_\text{rad} = \frac{3\kappa L P}{64\pi \sigma_\text{SB}G M T^4} \label{eq:rad gradient}
			\end{equation}
			depending on the luminosity $L$ of the planet and the envelope opacity $\kappa$. Following \citet{mordasiniGrainOpacityBulk2014}, we reduce the full interstellar opacity \citep{bellUsingFUOrionis1994} by a factor of 0.003. This value is a fit to detailed simulations of the grain dynamics in protoplanetary atmospheres \citep{movshovitzOpacityGrainsProtoplanetary2008,movshovitzFormationJupiterUsing2010}. The total luminosity includes the energy contribution due to the accretion of solids and gas, as well as the contraction of the envelope \citep{mordasiniCharacterizationExoplanetsTheir2012,mordasiniCharacterizationExoplanetsTheir2012a,alibertTheoreticalModelsPlanetary2013}. Solving the structure equations is crucially important to self-consistently account for the feedback of planetary luminosity and gas accretion.\\
			The accreted gas mass is determined iteratively by comparing the envelope masses between two iterations \citep{alibertModelsGiantPlanet2005}. In the beginning, the gas accretion is limited by the capacity of the planet to cool given its luminosity. As the core mass of the planet increases, the cooling can be so efficient that the gas accretion is limited by the supply of gas from the disk. Once the planet reaches this threshold, the planet is considered to be detached from the surrounding gas disk, accreting gas at the disk-limited gas accretion rate following \citet{bodenheimerDEUTERIUMBURNINGMASSIVE2013}. In past iterations of this and similar models, the disk limited gas accretion rate was either constrained by the radial flow of the gas or used a Bondi- or Hill-like accretion scheme \citepalias{emsenhuberNewGenerationPlanetary2021}. Both are inconsistent with the expected reduction of gas accretion due to the formation of a gap. This effect could only be ignored assuming eccentric orbits where the planet can efficiently access disk material despite the gap which is not applicable to the circular case of a single forming planet \citep{lubowDiskAccretionHighMass1999,brydenTidallyInducedGap1999}. In \citet{bodenheimerDEUTERIUMBURNINGMASSIVE2013}, this reduction to the gas accretion rate is taken into account.
   
		\subsection{Orbital migration} \label{sec:orbital migration}
		
		    A growing planet excites density waves in the gas disk through the inner- and outer Lindblad resonances as well as the corotation resonances \citep{goldreichExcitationDensityWaves1979, korycanskyNumericalCalculationsLinear1993}. A net torque is exerted on the planet resulting in orbital migration, so-called \emph{type-I} migration \citep{wardSurvivalPlanetarySystems1997, tanakaThreeDimensionalInteractionPlanet2002}. The positive torque of the Lindblad resonances inside the planetary orbit and the negative torque of the outer resonances usually result in migration towards the star. The corotation torque can be positive or negative, allowing outward migration for lower mass planets \citep{dittkristImpactsPlanetMigration2014}. The net torque depends on the local gas surface density gradient, the temperature profile, and the entropy. We follow the approach of \citet{colemanFormationPlanetarySystems2014} based on the torques of \citet{paardekooperTorqueFormulaNonisothermal2011} including the attenuation of the corotation torque due to eccentricity and inclination \citep{bitschOrbitalEvolutionEccentric2010, fendykeCorotationTorqueLowmass2014, colemanFormationPlanetarySystems2014}.\\
		    
		    As the planet grows more massive, it tidally interacts with the gas disk, locally decreasing the gas surface density until a gap forms \citep{linTidalInteractionProtoplanets1986}. In this so-called \emph{type-II} migration regime, the orbital migration rate can be significantly lower than in the type-I regime. We use the gap opening criterion of \citet{cridaWidthShapeGaps2006} as the transition threshold from type-I to type-II migration for a planet of mass $M$ orbiting with semi-major axis $a$
		    \begin{equation}
		        \frac{3H}{4 R_H} + \frac{50 \nu M_\star}{M a^2\Omega_K} \leq 1. \label{eq:migration transition criterion}
		    \end{equation}
            We adopt the smooth transition from the type-I to the type-II regime of \citet{dittkristImpactsPlanetMigration2014} and for the type-II migration direction and rate, we follow their approach where the planet moves along with the radial velocity of the gas \citep{pringleAccretionDiscsAstrophysics1981}. For even higher mass planets, the migration rate is limited by the disk-to-planet mass ratio corresponding to the fully suppressed case in \citet{alexanderGIANTPLANETMIGRATION2009}.
			
		\subsection{Pebble formation model} \label{sec:pebble formation model}
		
			 The dust surface density $\Sigma_\text{dust}=Z_\text{dust} \, \Sigma_\text{gas}$ follows the evolution of the gas disk surface density. The dust disk provides the mass reservoir for pebble formation and we identify the fraction $\text{f}_\text{peb} = Z_\text{dust} / Z_\text{tot}$ as the initial dust fraction. Since it is the parameter that is varied to change the amount of pebbles in the disk, we call it the pebble fraction. It marks the theoretical maximum fraction of solids that can be converted into pebbles. Because the dust surface density decreases with time following the gas surface density evolution, not all of the initial dust is converted into pebbles. Given the total solids-to-gas ratio $Z_\text{tot}$ and a fixed value $0 \leq \text{f}_\text{peb} \leq 1$, which are both initial conditions of the model, the planetesimals-to-gas fraction is simply $Z_\text{plan}=Z_\text{tot}(1-\text{f}_\text{peb})$. Note that in this way, the total initial solid mass in the system is independent of the value of f\textsubscript{peb}. However, since the different species of solids do not evolve in the same way, the available solid mass at a later time is strongly impacted by the choice of f\textsubscript{peb}. Most notably, pebbles neither form nor drift in the absence of the gas disk, whereas planetesimal accretion is not directly tied to the lifetime of the disk.\\
			
			 The location of pebble formation $r_g$, called growth radius, is defined by equating the pebble formation and drift time scales. Assuming Epstein drag, \citet{lambrechtsFormingCoresGiant2014} find
				\begin{equation}
					r_g(t) = \left( \frac{3}{16} G M_\star\right)^{1/3} \left( \epsilon_d Z_\text{dust} \,t \right)^{2/3}. \label{eq:rg}
				\end{equation}
			Here, $\epsilon_d=0.5$ is a free dust to pebble growth parameter. We use this prescription to determine the location of pebble formation given the initial dust-to-gas ratio $Z_\text{dust}$. The outward moving growth radius leaves behind inward drifting pebbles, inducing a mass flux
				\begin{equation}
					\dot{M}_\text{peb}(r) = 2 \pi r_g \frac{d r_g}{dt} \Sigma_\text{dust}(r_g) \label{eq:flux}
				\end{equation}
			for $r<r_g$. Equation \eqref{eq:flux} assumes that the pebble flux instantaneously adapts to the conditions at $r_g$. The pebble surface density inside the growth radius is then given by
				\begin{equation}
					\Sigma_\text{peb} = \frac{\dot{M}_\text{peb}}{2 \pi r v_r},
				\end{equation}
			assuming all of the dust converts to pebbles. This means that $\Sigma_\text{dust}$ vanishes and is replaced by $\Sigma_\text{peb}$ inside of $r_g$. Pebbles drift radially with the velocity $v_r$, depending on the Stokes number St \citep{weidenschillingAerodynamicsSolidBodies1977}
				\begin{equation}
					v_r = -2\frac{\text{St}}{\text{St}^2 + 1} \Delta v,
				\end{equation}							
			 where $\Delta v = \eta v_K$ is the sub-Keplerian headwind velocity given by the Kepler velocity $v_K = r \Omega_K$ and  $\eta = -\frac{1}{2}\left( \frac{H}{r}\right)^2 \frac{\partial \ln P}{\partial \ln r}$ for a disk scale height $H$ and a pressure $P$ at a radius $r$. \citet{lambrechtsFormingCoresGiant2014} find a typical pebble Stokes number of
				\begin{equation}
					\text{St} \approx \frac{\sqrt{3}\epsilon_p Z_\text{dust}}{8 \eta}, \label{eq:St}
				\end{equation}
			with a pebble growth efficiency $\epsilon_p = 0.5$. We adopt this prescription outside the ice line. We ignore erosive collisions of pebbles for both the pebble formation timescale as well as the resulting pebble size. By assuming pebble growth is only limited by radial drift, the pebble sizes are slightly overestimated in the inner parts of the disk at early times. In more turbulent disks, depending on the fragmentation velocity, the fragmentation of pebbles can be non-negligible resulting in different pebble sizes \citep{birnstielGasDustEvolution2010}. The pebble size and size distributions affect the disk opacity and in turn the disk structure \citep{savvidouInfluenceGrainGrowth2020}. Smaller pebbles drift more slowly resulting in higher pebble surface densities. The pebble flux however, does not change in this model as it is directly given by the radial velocity of the growth radius.\\
			The abundance of the dominant volatile species in icy pebbles (only water in this model) is assumed constant over the course of their inward drift up to the ice line in accordance with \citet{eistrupChemicalEvolutionIces2022}. After the sublimation of ice at the ice line crossing \citep{idaFormationDustrichPlanetesimals2016}, the growth and drift time scales do not balance any more and equation \eqref{eq:St} does not hold in the rocky pebble region. For the approximately chondrule sized \citep{morbidelliGreatDichotomySolar2015,shibaikeGalileanSatellitesFormed2019} rocky pebbles inside the ice line we use the definition of the Stokes number \citep{weidenschillingAerodynamicsSolidBodies1977}
				\begin{equation}
					\text{St} = \frac{t_\text{stop}\, v_K}{r}
				\end{equation}
			where $t_\text{stop}$ is the stopping time due to the gas drag, depending on the particle size and the local gas properties. The Stokes number is calculated in the appropriate drag regime \citep{rafikovFastAccretionSmall2004} assuming a particle radius of 1 mm \citep{friedrichChondruleSizeRelated2015}. We model the consequential pebble mass loss with a reduction of the pebble mass flux by a factor of 0.5 inside the ice line.\\
			
			At $t=0$, the solids in the disk consist of a planetary embryo, planetesimals, and dust which is forming pebbles. If planetesimals and embryos form from pebble-like objects themselves, we slightly overestimate planetary growth in the first $\sim10^5$ years by using this approach. There are models that connect the formation of planetesimals to the pebble disk, for instance by using a pebble flux-regulated approach and invoking the streaming instability in local pebble traps \citep{lenzPlanetesimalPopulationSynthesis2019,voelkelEffectPebbleFluxregulated2020}. In \citet{voelkelLinkingPlanetaryEmbryo2021a, voelkelLinkingPlanetaryEmbryo2021}, this approach is extended to the dynamic formation of embryos from the formed planetesimals. However, since the dust is quickly converted into pebbles in anywhere between $10^5 - 10^6$ years depending on the disk and since planetesimal accretion onto $10^{-2}M_E$ objects is inefficient, the planetary embryo can not grow significantly by planetesimal accretion in the time needed until the dust is converted into pebbles. This means that the disk quickly consists of planetesimals and a planetary embryo that is mainly accreting pebbles. Hence, the overestimation stemming from using this simplified approach is small and does not impede the main goal of understanding the interplay of the two accretion mechanisms.
   
		\subsection{Pebble accretion model} \label{sec:pebble accretion model}
		
			We consider the accretion of pebbles onto planets following \citet{johansenFormingPlanetsPebble2017}. The relative velocity $\delta v$ of a pebble approaching a protoplanet is given by
				\begin{equation}
					\delta v = \Delta v + \Omega_K R_\text{acc},
				\end{equation}				 
			where $R_\text{acc}$ is the accretion radius of a planet as defined below. For lower mass planets, the pebble approach velocity is dominated by the headwind $\Delta v$ compared to the Keplerian motion of the planet. This is referred to as the headwind or Bondi regime. For more massive planets, $\delta v$ is dominated by the shear velocity. We consider this to be the case when the Hill speed $v_H = \Omega_K R_H$ exceeds the headwind velocity $\Delta v$, entering the shear or Hill regime. This represents a transition as the planet reaches the mass $M = 3  \eta^3 M_\star$. In the strong pebble-protoplanet coupling limit, where friction time scales are short compared to encounter time scales, the accretion radii in the headwind regime (top) and the shear regime (bottom) are given by \citep{johansenFormingPlanetsPebble2017}
				\begin{equation}
					R'_\text{acc} =
                	\begin{cases}
                	 \left( \frac{4\, \tau_f \Delta v}{R_B} \right)^{1/2} R_B \; , \\
                	 \\
                	 \left( \frac{\Omega_K \tau_f}{0.1} \right)^{1/3} R_H \;,
                	\end{cases}
				\end{equation}	
			with $\tau_f = \text{St}/\Omega_K$, the Bondi radius $R_B=\frac{GM}{\Delta v^2}$, and the Hill radius $R_H$. To account for weaker interactions when the friction time scale is longer
			 than the encounter time scale $t_e = G M / (\Delta v + \Omega_K R_H)^3$, the accretion radii are modified by \citep{ormelEffectGasDrag2010}
				\begin{equation}
					R_\text{acc} = R'_\text{acc} \; e^{-0.4\left(\tau_f / t_e\right)^{0.65}}.
				\end{equation}
			
			 We further distinguish between 3D accretion, where the accretion region is fully embedded in the pebble disk, and the more efficient 2D accretion, which occurs when the accretion radius $R_\text{acc}$ reaches beyond the pebble scale height $H_\text{peb} = H \left( 1 +\frac{\text{St}}{\alpha}\frac{1 + 2 \text{St}}{1+ \text{St}} \right)^{-1/2}$ \citep{youdinParticleStirringTurbulent2007}. Here $\alpha=0.002$ is the $\alpha$-viscosity parameter. The 2D and 3D pebble accretion rates are
				\begin{eqnarray}
					\dot{M}_\text{2D} & = & 2 R_\text{acc} \Sigma_\text{peb} \delta v,\label{eq:mdot_2d}\\
					\dot{M}_\text{3D} & = & \pi R^2_\text{acc} \rho_\text{peb} \delta v, \label{eq:mdot_3d}
				\end{eqnarray}
            where $\rho_\text{peb} = \Sigma_\text{peb} / (\sqrt{2 \pi}H_\text{peb})$ is the midplane pebble density.
            Inserting the appropriate expression for $R_\text{acc}$ into equations \eqref{eq:mdot_2d} and \eqref{eq:mdot_3d} respectively, yields four possible pebble accretion rates.\\
            
            Pebble accretion stops when the pebble flux vanishes. This can be due to the exhaustion of the outside solid mass reservoir. In this model, this corresponds to the growth radius $r_g$ reaching the outer edge of the gas disk. Another mechanism for stopping the pebble flux is the so-called pebble isolation mass \citep{lambrechtsFormingCoresGiant2014} \citep[see also][]{ataieeHowMuchDoes2018,bitschPebbleIsolationMass2018, shibaikePlanetesimalFormationGas2020}
            \begin{equation}
                \text{M}_{\text{iso}} \approx 20 \left( \frac{H/r}{0.05}\right)^3 \text{M}_\text{E}. \label{eq:M_iso}
            \end{equation}
		    At this mass, the planet perturbs the gas disk outside the planet sufficiently in order to create a region of super-Keplerian gas flow. In this zone, the drifting pebbles from further outside encounter a tailwind instead of a headwind. Thus, pebbles stop drifting and pile up outside the planet. This stops the pebble accretion onto the planet responsible for this pressure bump, as well as starving all potential inside planets of pebbles \citep{paardekooperDustFlowGas2006}. The value of the pebble isolation mass depends on the particular disk via the scale height $H$ in this prescription but, typically, it is equal to roughly one Earth mass at 0.1 AU and increases to $20-30$ M\textsubscript{E} at 1 AU due to the flared disk structure. Beyond that distance, planet cores almost never reach the even larger pebble isolation mass because the growth radius reaches the outer disk edge before. This does not imply that core masses do not easily exceed tens of Earth masses outside of 1 AU when pebble accretion stops. Even though planets are exposed to the flux of pebbles for a shorter amount of time, depending on the disk, core growth can be significant up to the maximum of 40 AU considered here.
		    
	\section{Population synthesis outcomes} \label{sec:population synthesis outcomes}

        \begin{table}
        \caption{Distributions of varied initial parameters of the population synthesis.}
        \label{tab:initial conditions}
            \centering
            \begin{tabular}{ccc}
                \hline\hline
                 Parameters & Mean & Deviation\\
                 \hline
                 $Z_\textsubscript{tot}$ & $\mu= -0.02$ & $\sigma=0.22$\\
                 $M_\textsubscript{gas}$ & $\log_{10}(\mu/\text{M}_\text{sol})=-1.49$ & $\sigma=0.35$ dex\\
                 $P_\star$ & $\log_{10}(\mu/d) = 0.676$ & $\sigma = 0.306$ dex\\
                 $\dot{M}\textsubscript{wind}$ & $\log_{10}(\mu/(\text{M}_\text{sol}\text{yr}^{-1}))=-4.7$ & $\sigma=1$ dex\\
            \end{tabular} 
        \end{table}
	
	    \begin{figure*}
            \centering
            \includegraphics[width=\linewidth]{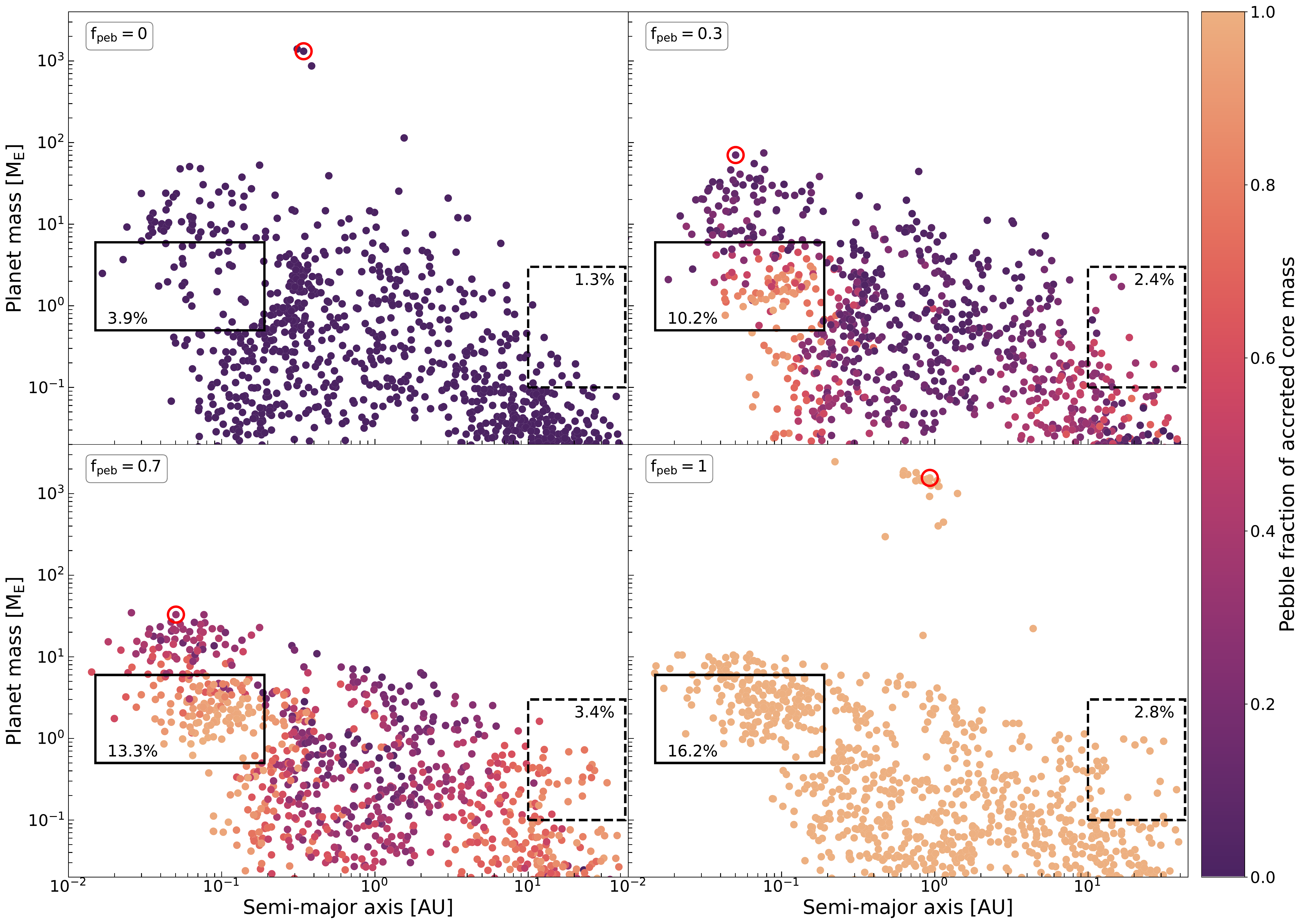}
                \caption{Planet mass over semi-major axis of one thousand single-planet simulations after 2 Gyr for pebble fractions f\textsubscript{peb} = 0 (\emph{planetesimals-only}), f\textsubscript{peb} = 0.3 (\emph{pebble-poor}), f\textsubscript{peb} = 0.7 (\emph{pebble-rich}), and f\textsubscript{peb} = 1 (\emph{pebbles-only}). The solid-line boxes highlight planet masses of 0.5 to 6 M\textsubscript{E} in the inner disk region up to 0.2 AU. The dashed-line boxes highlight planet masses above 0.1 M\textsubscript{E} outside of 10 AU. The boxes are labelled with the percentage of planets in these regions. The colour of the points indicates the fraction of accreted pebbles compared to the total mass of accreted solids. The darkest points are fully planetesimal-formed planets and the brightest points are planets formed only by pebbles. The encircled points are planets that formed from the same disk with different pebble fractions. Their formation paths are further examined in Sect. \ref{sec:giant planet formation}}
                \label{fig:mvsa_ring}
            \end{figure*}
	    
		We simulate the formation and evolution of 1000 single-planet systems around solar mass stars for different fixed values of f\textsubscript{peb}. Since we focus on the formation stage rather than the long term evolution of planets, we present populations after 2 Gyr of time evolution which is well beyond the longest gas disk lifetimes of up to $10^7$ years considered here. We compare the planetesimals-only case, where f\textsubscript{peb} = 0, to the pebble-poor (f\textsubscript{peb} = 0.3), the pebble-rich (f\textsubscript{peb} = 0.7), and the pebbles-only (f\textsubscript{peb} = 1) cases. Following \citetalias{emsenhuberNewGenerationPlanetary2021}, we use planetesimals with a diameter of 600 meters, a fixed viscosity $\alpha = 0.002$, and an initial gas disk slope parameter of $\beta = 0.9$ for all populations. For each of the $10^3$ systems within a population, the initial gas disk mass, the inner radius $R\textsubscript{in}$, and the total solids-to-gas ratio $Z\textsubscript{tot}$ are varied. In order to compare the populations using different pebble fractions, we choose the same initial conditions for all four sets of simulations.\\
        The solids-to-gas ratio $Z_\text{tot}$ is given by normally distributed stellar metallicities of \citet{santosSpectroscopicMetallicitiesPlanethost2005} under the assumption of an equal disk and stellar dust-to-gas ratio. We use the gas disk masses of \citet{tychoniecVLANascentDisk2018} which are obtained from continuum dust emission spectra assuming a dust-to-gas ratio of 0.01. We assume a log-normal distribution and correct for the actual, not necessarily equal to 0.01, solids-to-gas ratio in our setup. The inner radius $R\textsubscript{in}$ is given by the corotation radius with respect to the stellar rotation which is obtained from a log-normal distribution of stellar rotation periods of T-Tauri stars \citep{venutiCSI2264Investigating2017}. Given these parameters, the characteristic radius $R\textsubscript{char}$ and the initial surface density at 5.2 AU $\Sigma_0$ are determined. The external photo-evaporation rate parameter $\dot{M}\textsubscript{wind}$ \citepalias[see][]{emsenhuberNewGenerationPlanetary2021} is also varied for each system following a log-normal distribution. Note that since the initial stellar mass is fixed to one solar mass, the initial internal photo-evaporation rate is not varied. In table \ref{tab:initial conditions}, we list the distribution parameters of the varied quantities. The distributions of the initial gas disk masses and the characteristic disk sizes are shown in Figs. \ref{fig:disks} and \ref{fig:rchar}. Note that, compared to \citetalias{emsenhuberNewGenerationPlanetary2021a}, the protoplanetary disks are slightly less massive since we now correct the assumed dust-to-gas ratio of 0.01 in \citet{tychoniecVLANascentDisk2018} to the actual one in the calculation of the initial gas disk mass.  In every disk, a 0.01 M\textsubscript{E} embryo is randomly placed at up to 40 AU following a log-uniform distribution at the beginning of the simulation.\\		
			
        The planet masses are displayed in Fig. \ref{fig:mvsa_ring} as a function of semi-major axis for the different populations. The colour shows the constitution of the accreted solid material: the darkest dots have accreted planetesimals only, whereas the lightest dots are dominated by pebble accretion.\\
        The top-left panel shows the synthesis outcome using only planetesimals without any pebbles present. It features a few giant planets above 100 M\textsubscript{E} around 0.3 AU. The fact that only a few giants form is due to the rather low-mass disks generated here as well as the absence of other planetary embryos \citepalias{emsenhuberNewGenerationPlanetary2021}. Nevertheless, this confirms once more that in disks that are massive enough and contain enough small planetesimals, it is possible to form giant planets. Around 0.1 AU, there is a lower number density of roughly Earth mass planets compared to the simulations containing increasing amounts of pebbles, shown in the top-right, bottom-left, and bottom-right panel (see solid-line boxes). This is explained by the fact that inner planets have access to much more mass in the form of drifting pebbles from the whole disk rather than locally available planetesimals. The more massive planets found in the same region are formed further outside, around a few AU, where growth via planetesimal accretion is efficient. These planets start to migrate inwards more quickly once they reach a few Earth masses (see Sect. \ref{sec:orbital migration}) populating the higher-mass demographic of the inner disk. Outside of 10 AU, there are few planets above 0.1 M\textsubscript{E} as shown by dashed-line box in the top-left panel. Low planetesimal accretion rates of low-mass planets in the outer regions of the disk, even with small 600 meter planetesimals, are expected \citepalias{emsenhuberNewGenerationPlanetary2021}. This is due to the low collision probability, large orbital period, and low planetesimal surface density in the outer disk.\\
        In the pebble-poor and pebble-rich populations shown in the top-right and bottom-left panel of Fig. \ref{fig:mvsa_ring}, no giant planets are formed. More precisely, there are no planets where the envelope mass exceeds the core mass and the maximal planet mass decreases to about 74 M\textsubscript{E} (f\textsubscript{peb} = 0.3) and 34 M\textsubscript{E} (f\textsubscript{peb} = 0.7) as the pebble fraction increases. More planets above a few Earth masses end up on close orbits compared to the planetesimals-only simulation. While roughly 24\% of planets grow more massive than one Earth mass in the planetesimals-only case, this percentage increases up to about 34\% with increasing pebble fraction. We find that, as a consequence, in the planetesimals-only case 15\% of all planets migrate to closer than half their initial distance whereas almost 23\% of all planets do so in the pebble-rich simulation. The increased number of strongly migrating planets is, however, not only due to the larger number of planets above one Earth mass. Due to the early growth by pebbles, planets are more massive while still inside a more dense gas disk which enhances migration rates. We find that among the planets that grow beyond one Earth mass, 59\% migrate significantly (decay more than half their initial separation) in the planetesimals-only case whereas 70\% do so in the pebbles-only case. Such increased migration rates for pebble-formed planets have been reported before \citep[e.g.][]{bruggerPebblesPlanetesimalsOutcomes2020}. Planets in this mass range normally enter the type-II migration regime due to runaway gas accretion, once pebble accretion stops. But since these planets do not end up rapidly accreting gas due to the continuously heated envelope by planetesimal accretion in our simulations, slower type-II migration is never reached. We investigate the (non-)formation of giant planets more closely in Sect. \ref{sec:giant planet formation}. Compared to the planetesimals-only population, we observe more planets approaching one M\textsubscript{E} outside of 10 AU (dahsed-line boxes) as well as more planets around a few Earth masses in the inner disk regions (solid-line boxes). These planets are increasingly pebble dominated with larger pebble fractions, as can be seen from the colour mapping in Fig. \ref{fig:mvsa_ring}. In the regions where growth via planetesimal accretion is efficient, many planets still end up accreting more of their mass in the form of planetesimals. This is possible since after pebble accretion stops, by reaching the pebble isolation mass, depletion of pebbles, or due to the dispersal of the gas disk, the planets continue to accrete planetesimals.\\
        In the pebbles-only simulations shown in the bottom-right panel, the before mentioned increased inward migration trend persists. Planets that do not accrete a large envelope never grow more massive than about 10 M\textsubscript{E}. However, some of the most massive planets can accumulate a large envelope and slow their migration significantly. Several giant planets of roughly one Jupiter mass and more are formed around and inside of 1 AU. This agrees with the previous findings of more frequent giant formation in pebble accretion models \citep{lambrechtsRapidGrowthGasgiant2012,bitschGrowthPlanetsPebble2015}. In the outer disk, planets grow more massive with increasing pebble fractions (see dashed-line box) since planetesimal accretion rates are low in this region. Pebbles on the other hand, can also be accreted at large distances once the growth radius moves past the planet. For f\textsubscript{peb}=1, the number density of planets above 0.1 M\textsubscript{E} in the outer region (dashed-line box) is again lower compared to the pebble-rich case because planets tend to migrate inside of 10 AU. The pebble accretion period ends when all the dust is converted to pebbles, ultimately limiting the core masses that can be reached in the pebbles-only scenario. 

	\section{Giant planet formation} \label{sec:giant planet formation}
	    
	    \begin{figure}
        \centering
        \includegraphics[width=\linewidth]{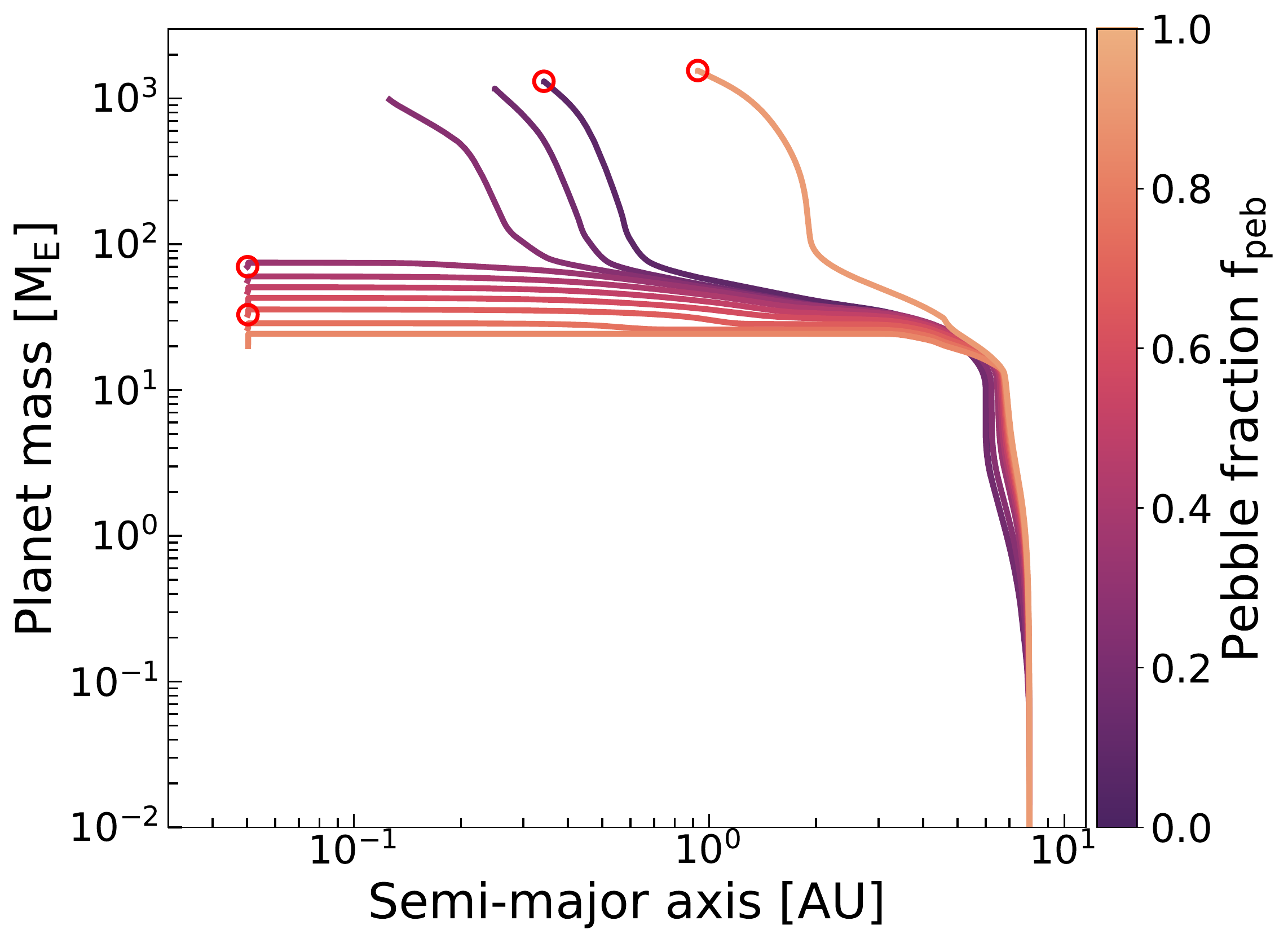}
            \caption{
                Formation tracks of a planet during 2 Gyr with pebble fractions f\textsubscript{peb} from 0 to 1 in increments of 0.1 (colours) using the same disk that gives rise to the encircled planets in Fig. \ref{fig:mvsa_ring}. The tracks of the four cases (f\textsubscript{peb} = 0, 0.3, 0.7, 1) shown in Fig. \ref{fig:mvsa_ring} are again marked by a red circle.
                }
            \label{fig:847_track}
        \end{figure}

   		\begin{figure}
        \centering
        \includegraphics[width=\linewidth]{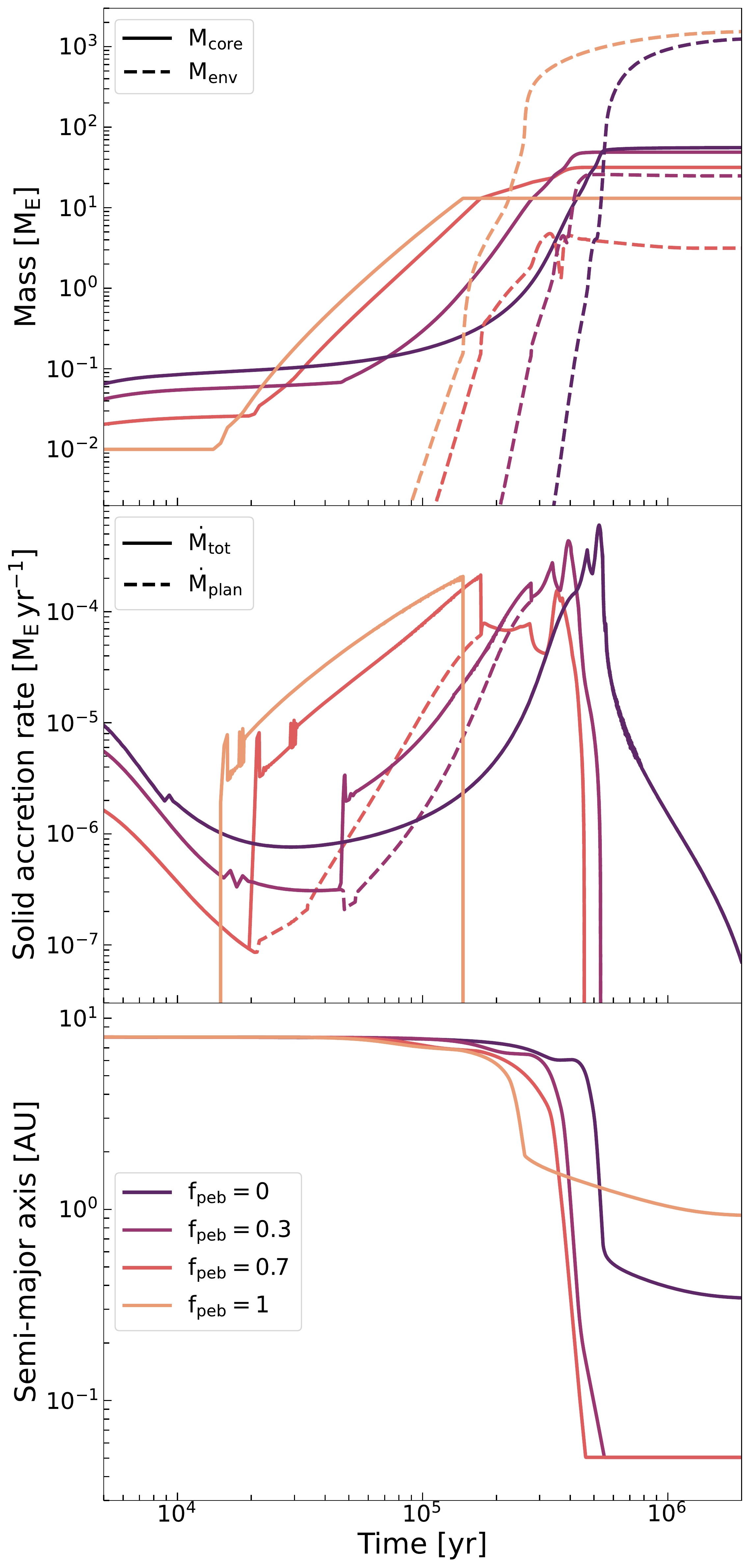}
            \caption{
                Time evolution of a planet forming in disks of varying pebble fraction f\textsubscript{peb} (the four encircled cases in Fig. \ref{fig:847_track}). The top panel shows the the core mass (solid lines) and the envelope mass (dashed lines), the middle panel shows the total solid accretion rate (solid lines) and the planetesimal accretion rate for the f\textsubscript{peb} = 0.3 and f\textsubscript{peb} = 0.7 cases (dashed lines). In the bottom panel, the semi-major axis over time is displayed.
                }
            \label{fig:847}
        \end{figure}
        
        \begin{table}
        \caption{Specific initial parameters of the system of interest in Sect. \ref{sec:giant planet formation}.}
        \label{tab:initial conditions 847}
            \centering
            \begin{tabular}{cc}
                \hline\hline
                 System specific parameters & Values \\
                 \hline
                 $Z_\textsubscript{tot}$ &   0.012 \\
                 $\Sigma_0$ &  237 g cm$^{-2}$ \\
                 $R_\text{in}$ & 0.049 AU \\
                 $R_\text{char}$ & 120.8 AU \\
                 $\dot{M}_\text{wind}$ & $9.1921\cdot10^{-6} \; \text{M}_\text{sol}\text{yr}^{-1}$\\
                 $a_\text{init}$ & 7.97 AU
            \end{tabular}

        \end{table}
    
        It is not surprising that giants can form from pebbles alone \citep{lambrechtsRapidGrowthGasgiant2012} or from small planetesimals alone \citepalias{emsenhuberNewGenerationPlanetary2021}. We do not aim to discuss giant formation pathways in those cases in detail again but use them as a reference for the hybrid setups. We focus on the mechanisms preventing giant formation that arise from the interplay of planetesimal and pebble accretion which can be best understood from the formation history on system-level.\\
        The formation of an envelope due to the accretion of gas is strongly coupled to the accretion of solids. On one hand, the increase of the core mass of a planet positively affects the onset of gas accretion. On the other hand, the liberated gravitational energy of the impacting solids heats the envelope, increasing the pressure, which is counteracting the pull of the planet on the surrounding gas. In addition, the different solid accretion rates due to pebbles or planetesimals strongly impact the migration behaviour of the planet. \\
	    
		Figure \ref{fig:847_track} shows an example where a giant planet is formed in the strongly planetesimal dominated cases with f\textsubscript{peb} = 0, 0.1, and 0.2.  As seen before, a giant planet can also form in the pebbles-only simulation (f\textsubscript{peb} = 1). In all other cases shown in pebble fraction increments of  0.1, no giant planet is formed. The formation paths of the encircled planets in Fig. \ref{fig:mvsa_ring}, corresponding to f\textsubscript{peb} = 0, 0.3, 0.7, 1, are again highlighted with a red circle at the planet mass and location at 2 Gyr. This disk is ideal in order to dissect the differences causing the strongly contrasting formation outcomes for different values of f\textsubscript{peb}. We note however, that the effects observed here are general since a similar pattern is observed in other systems that form giant planets in the pebbles-only case but fail to produce giants when a fraction of the solids is in the planetesimals. We hence consider it a representative example when it comes to giant (non-)formation in our simulations. The initial conditions of this particularly giant planet favouring disk are shown in table \ref{tab:initial conditions 847}. Listed are the total solid-to-gas ratio $Z_\textsubscript{tot}$, the gas surface density at 5.2 AU $\Sigma_0$, the inner and characteristic disk radii $R_\text{in}$ and $R_\text{char}$, the external photo-evaporation parameter $\dot{M}_\text{wind}$, and the initial position of the embryo $a_\text{init}$.\\
		After an initial phase of inward migration from its starting location at almost 8 AU, the planet can migrate outwards slightly before significantly migrating inwards in all simulations. The planetesimals-only planet (darkest line) grows massive enough to trigger runaway gas accretion, carving a gap in the gas disk and subsequently migrating slower in the type-II migration regime. The same happens in the 10\% and 20\% pebble fraction cases but the inward migration is stronger, causing the runaway gas accretion to happen when the planet is already closer in. This is explained by the increased early core growth rate due to the accretion of pebbles. The outcome is a planet that ends up on a closer in orbit the higher the pebble fraction is. The pebbles-only planet (brightest line), on the other hand, grows so fast that it reaches a higher core mass more quickly, entering type-II migration earlier and on a wider orbit. At 2 Gyr, the mass of the formed giant lies between 3 and 4.9 M\textsubscript{J} and orbits between 0.15 and 0.9 AU. In all the other simulations with pebbles and planetesimals in the disk, the planet migrates all the way to the inner disk edge at about 0.05 AU and has a mass between 23 and 70 M\textsubscript{E}. Note that they lose a small amount of envelope mass over time due to photo-evaporation close to the star.\\
		It is apparent from the tracks shown in Fig. \ref{fig:847_track} that even when only a small fraction of the mass is in the planetesimals, the formation of giant planets is suppressed. We find that, in this particular disk, a pebble dominated giant planet can only form when the fraction of planetesimals is below 2\%, i.e. for f\textsubscript{peb} $>$ 0.98. We further note that, as mentioned already in Sect. \ref{sec:population synthesis outcomes}, the increase of the amount of pebbles with respect to planetesimals does not lead to a higher final mass of the planet. Rather, it leads to smaller final planetary masses in the case of these large planets that almost grow to giant planets.\\
		
		The formation pathway of the same system is again presented in Fig. \ref{fig:847} in terms of core and envelope mass, core accretion rate, and semi-major axis as a function of time. For the sake of clarity, we only show the simulations using the pebble fraction values f\textsubscript{peb} = 0, 0.3, 0.7, and 1. In all cases, the growth radius $r_g$ has not yet reached the embryo's location before $10^4$ years. In this early phase, only planetesimal accretion is possible and the total core accretion rate is equal to the planetesimal accretion rate (the dashed and solid lines in the middle panel overlap). Since the planetesimals are in an equilibrium state with respect to self-stirring at the initialisation of the simulation, the accretion rates can be moderate. Within a few $10^4$ years, the embryo starts exciting the planetesimal dynamical state and the planetesimal accretion rate drops as a result. Unsurprisingly, the rates are lower when less planetesimals are present in the disk.\\
		When $r_g$ moves outside of the planet orbit, there is an immediate increase in the total core accretion rate due to the onset of pebble accretion. This happens earlier for higher pebble fractions due to the $Z_\text{dust}^{2/3}$ dependence of the growth radius. As a result, planet cores are formed earlier the larger the pebble fraction is. The pebble accretion rate is higher for larger values of f\textsubscript{peb}. In disks containing more than 70\% pebbles, pebble accretion is always more dominant than planetesimal accretion for planets below the pebble isolation mass. In the pebble-poor case (f\textsubscript{peb}=0.3), the large amount of planetesimals allows for planetesimal accretion rates to become comparable to the accretion rate of pebbles once the core grows more massive. Note that the initial spike in the total solid accretion rate at the start of pebble accretion is a numerical effect due to the crossing of the growth radius and the planet location, which has an insignificant effect on the planet formation pathway and final outcome. The additional step-like features in the pebble-rich (f\textsubscript{peb}=0.7) and pebbles-only (f\textsubscript{peb}=1) cases are due to changes of the pebble accretion regime according to Sect. \ref{sec:pebble accretion model}.\\
		The planet stops accreting pebbles between 0.1 and 0.3 Myr depending on the pebble fraction, causing the visible drop in accretion rate (see all lines except the darkest). The value of the pebble isolation mass increases towards greater separations from the star given by \eqref{eq:M_iso}. At this point in time, the planet orbits at 7 AU in all simulations that contain pebbles. In this region, M\textsubscript{iso} is above the 13 M\textsubscript{E} of the planet in all cases. Hence, the pebble accretion stops because all the dust is converted into pebbles all the way to the outer disk edge. As noted before, the growth radius moves through the disk more rapidly the larger the dust fraction is which results in the pebble accretion phase ending earlier for higher pebble fractions.\\
		After pebble accretion stops, the core can only grow further through planetesimals for the rest of the formation process so the total core accretion rate is again equal to the planetesimal accretion rate. The availability of planetesimals, meaning the value of f\textsubscript{peb}, dictates the core accretion rate. In the f\textsubscript{peb} = 1 case, this means the final core mass is directly limited to the pebble isolation mass or by the mass reached when the pebble flux ceases.\\
		As shown in the top panel of Fig. \ref{fig:847}, the planet can already accumulate a small envelope during the pebble accretion phase. When the core accretion rate suddenly drops, the gas accretion rate increases immediately due to the lowered luminosity associated with solid accretion, resulting in a steep increase of the envelope mass. In the f\textsubscript{peb} = 1 case, the planet undergoes runaway accretion of gas as already seen in Fig. \ref{fig:847_track}. In both the hybrid cases however, the envelope mass never exceeds the core mass and the gas accretion is not sufficient to form a gap which would slow down the inward migration. As a consequence, the planet moves to the inner disk edge in just about 0.2 Myr.
		
		\subsection{Delay of runaway gas accretion} \label{sec:delay of runaway gas accretion}
		
    		\begin{figure}
    		\centering
            \includegraphics[width=\linewidth]{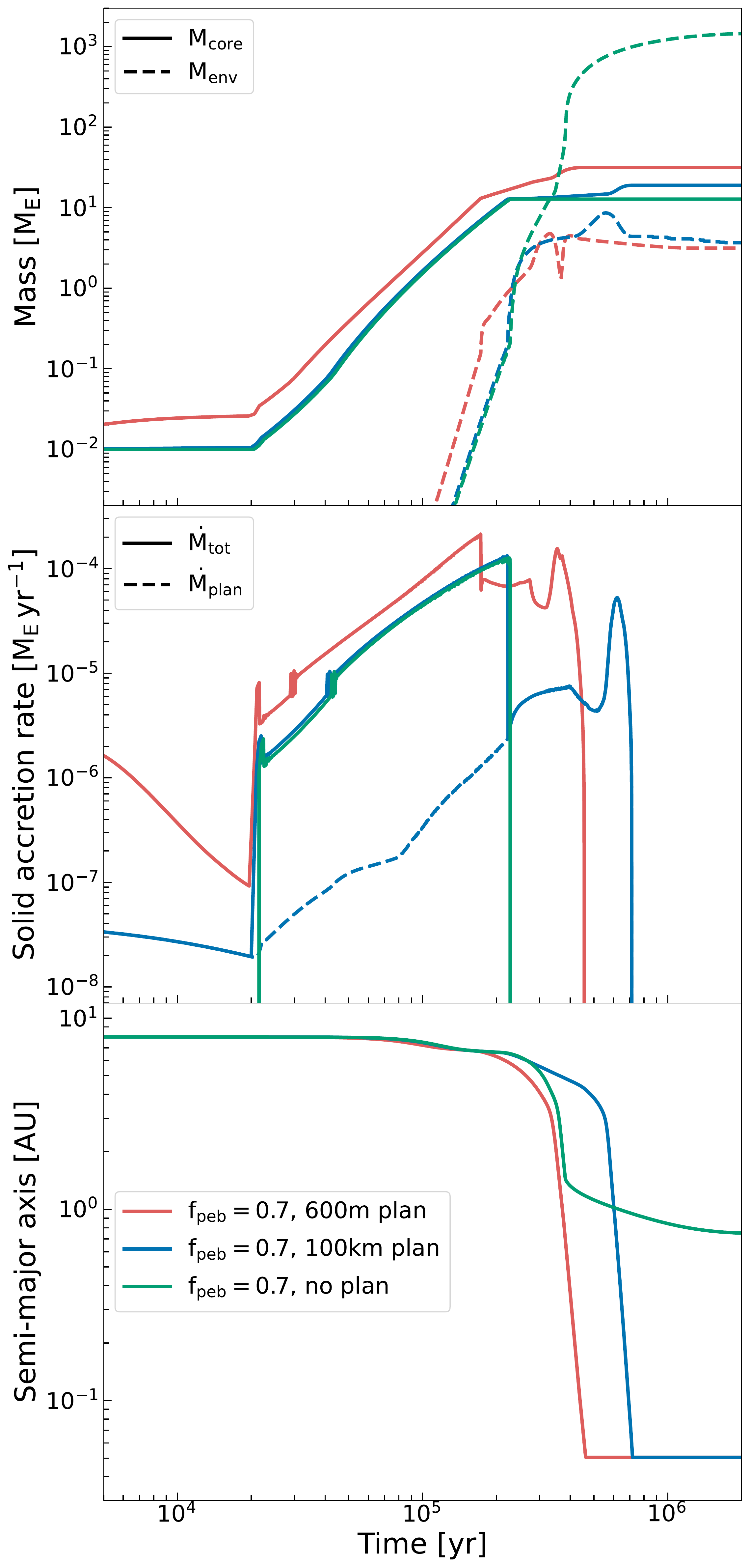}
                \caption{
                    Time evolution of a planet forming in a f\textsubscript{peb} = 0.7 disk with 100 km planetesimals (blue) and disabled planetesimal accretion (green). The nominal f\textsubscript{peb} = 0.7 case using 600 m planetesimals (red) is again shown for comparison. The top panel shows the core mass (solid lines) and the envelope mass (dashed lines), the middle panel shows the total core accretion rate (solid lines) and the planetesimal accretion rate (dashed line). In the bottom panel, the semi-major axis over time is displayed.
                    }
                \label{fig:847_noplan}
            \end{figure}
    		
    		The onset of rapid gas accretion is prevented when the planet keeps accreting planetesimals after pebble accretion stops. This suggests that the remaining accretion heating due to planetesimals is responsible for the delay of runaway gas accretion. It begs the question however, whether this finding is just a result of the small size of the planetesimals chosen. By using large 100 km diameter planetesimals, we test the f\textsubscript{peb} = 0.7 case for lower planetesimal accretion rates and subsequently less envelope heating. The blue lines in Fig. \ref{fig:847_noplan} show the equivalent formation pathway in the large planetesimal case as the red lines representing the 600 m simulation. The red lines, shown as a comparison, are identical to the ones in Fig. \ref{fig:847}. The general outcome remains the same but the planetesimal accretion rate is reduced by about an order of magnitude. Albeit lower, the heating is still sufficient to prevent runaway gas accretion as the envelope mass does not exceed the core mass and the planet still migrates inwards all the way through the disk. This is compatible with the minimum core accretion rate to prevent runaway gas accretion of roughly $10^{-5}$$-$$10^{-6}$ M\textsubscript{E}yr\textsuperscript{-1} predicted in \citet{alibertFormationJupiterHybrid2018}.\\
    		The green line in Fig. \ref{fig:847_noplan} shows the exact same setup but the accretion of planetesimals is disabled. Disregarding the low planetesimal accretion rates early on, these planets follow the same formation path up to the end of pebble accretion as in the blue case, even though there are less available solids in the disk. After pebble accretion stops, the core mass is fixed and no further heating due to the accretion of solids can occur. The planet enters the runaway gas accretion regime shortly after since the envelope cools rapidly in the absence of solid accretion. As before in the planetesimals-only and pebbles-only cases, the planet migrates more slowly in the type-II regime allowing them to halt outside the inner disk edge. Instead of moving all the way inside, as with ongoing planetesimal accretion, a 4.5 M\textsubscript{J} planet is formed at 0.75 AU.\\
    		
    		Since the remaining accretion of planetesimals after pebble accretion stops is the only difference between the green and blue curves, we identify the associated heating as a crucial mechanism that is preventing giant formation in hybrid pebble-planetesimal disks in our model. This has, however, also consequences on the migration of planets. We study the role of migration on giant formation in the following section.
		
		\subsection{Inward migration} \label{sec:inward migration}
		
    		\begin{figure}
    		\centering
            \includegraphics[width=\linewidth]{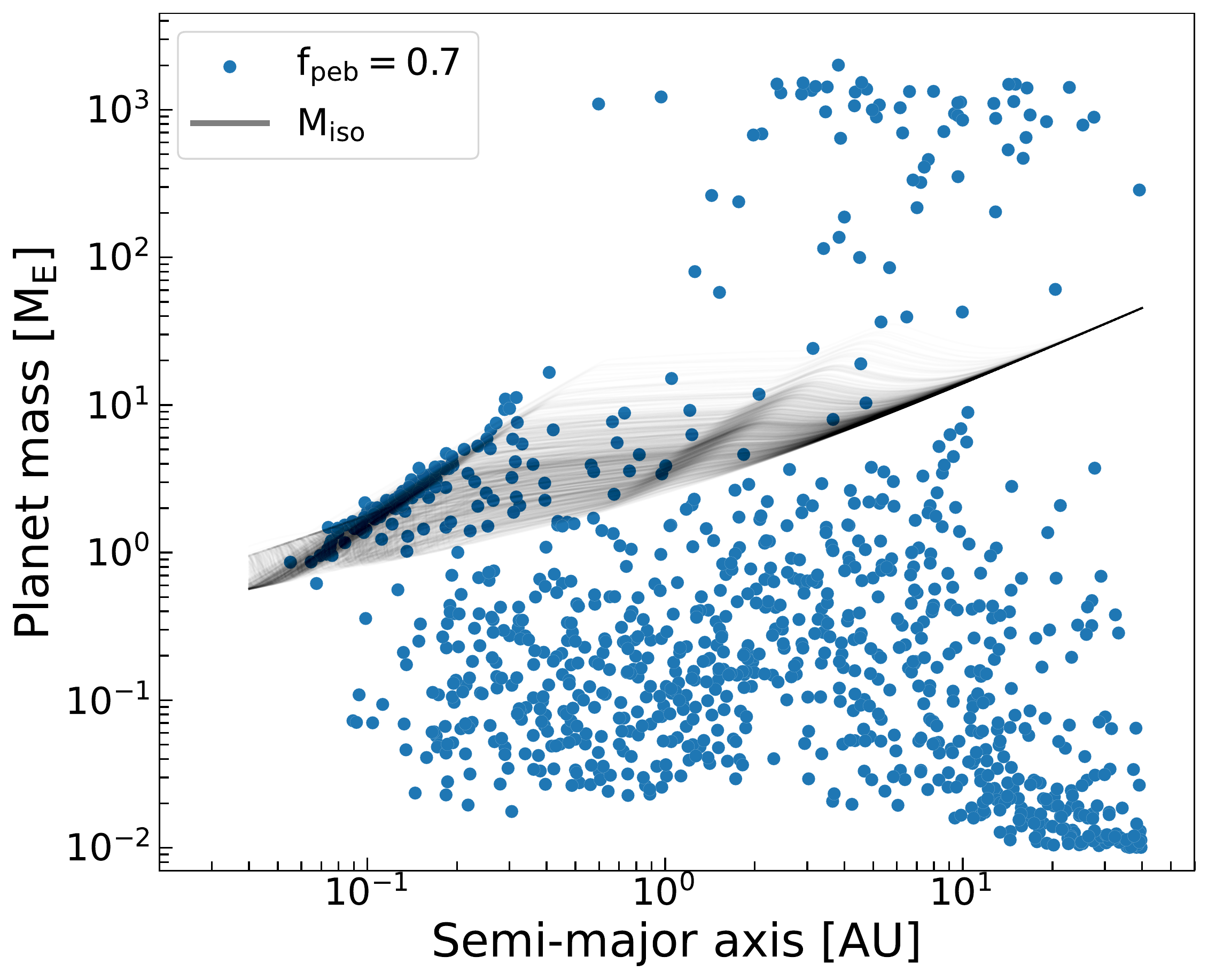}
                \caption{
                    Planet mass over semi-major axis diagram of the \emph{in-situ} population for f\textsubscript{peb} = 0.7. The opaque black lines are the radial pebble isolation mass profiles of all disks at $10^5$ years.
                    }
                \label{fig:0p7_nomig}
            \end{figure}
            
            \begin{figure}
    		\centering
            \includegraphics[width=\linewidth]{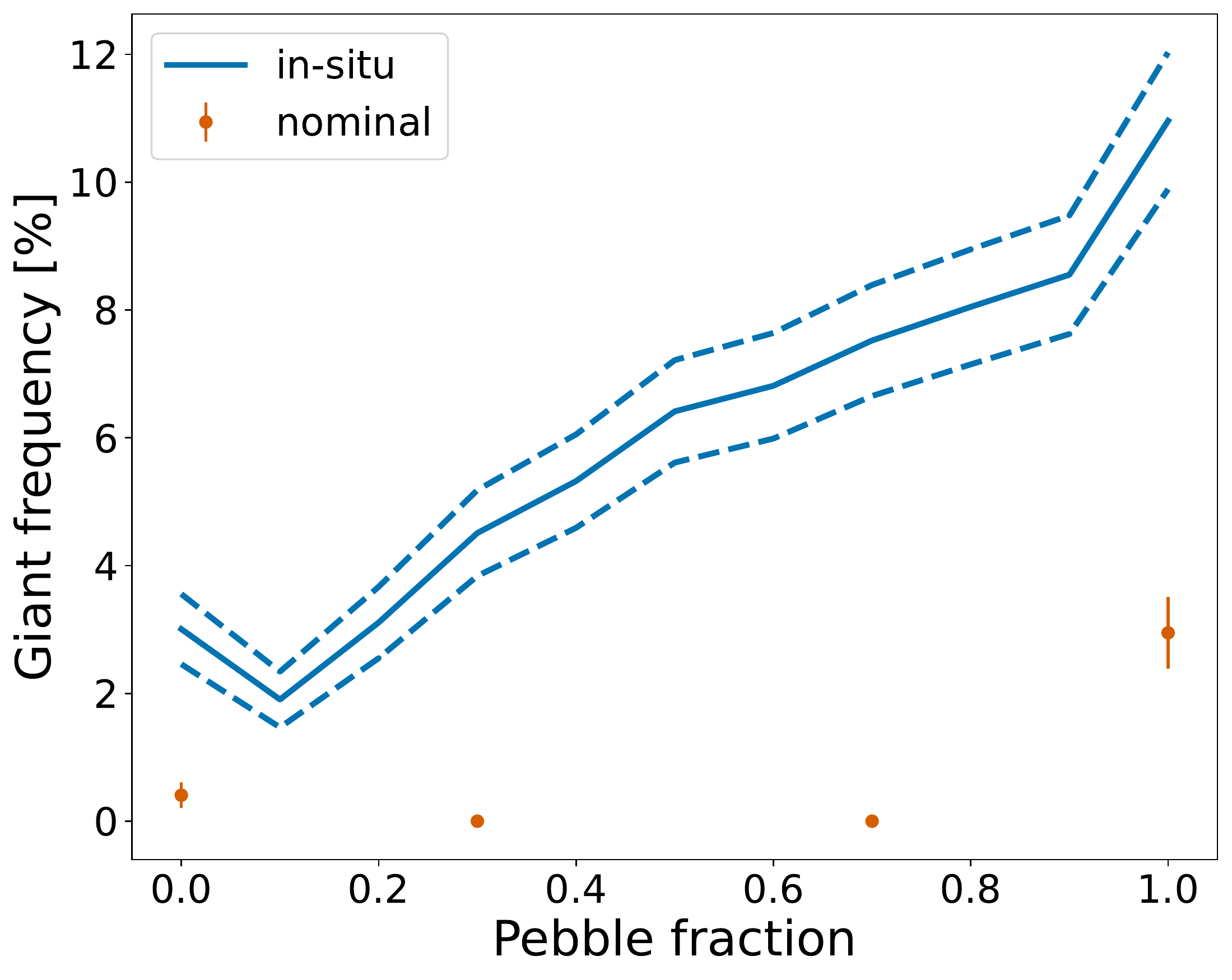}
                \caption{
                    Frequency of giants formed in the single-planet \emph{in-situ} simulations depending on the fraction of pebbles are shown in blue. For comparison, the giant planet frequencies obtained from the nominal populations in Fig. \ref{fig:mvsa_ring} are shown in orange.
                    }
                \label{fig:ngiants_freq}
            \end{figure}
    		
    		Since, after pebble accretion stops, the core keeps growing through planetesimal accretion and gas accretion is slowed but not halted entirely, the onset of runaway gas accretion is delayed and not necessarily impossible. The reason why giant formation is prevented altogether in our simulations, is because massive planets that are just about to cross the gas runaway threshold migrate to the inner disk edge within a few $10^5$ years (see bottom panels in Figs. \ref{fig:847} and \ref{fig:847_noplan}) before they can accrete gas rapidly and carve a gap. We contrast the nominal f\textsubscript{peb} = 0.7 population shown in Fig. \ref{fig:mvsa_ring} with the same population without migration to underline this. While the \emph{in-situ} formation of planets is unlikely, it can give a good impression of the impact migration has. As shown in Fig. \ref{fig:0p7_nomig}, there is an abundance of giant planets formed from 1 AU all the way to 40 AU in the f\textsubscript{peb} = 0.7 case without migration. Note that in the inner disk, the pebble isolation mass is too low to allow runaway gas accretion. As a consequence, there is an over-density of planets following the $(H/r)^3$ slope of the isolation mass prescription. These planets correspond to the pebble dominated planets in the inner disk which exist in every simulation with pebbles shown in Fig. \ref{fig:mvsa_ring} but when also considering migration, the pebble isolation mass slope is washed out in the mass over semi-axis diagram. Since the disk aspect ratios vary and evolve over time, the pebble isolation mass is different for all disks. The opaque black lines in Fig. \ref{fig:0p7_nomig} are the pebble isolation masses as a function of distance for all disks at $10^5$ years. They give an intuition for the value of the pebble isolation mass in the different disk regions at the time when inner planets typically approach this mass range.\\
            In Fig. \ref{fig:ngiants_freq}, the frequency of giants forming \emph{in-situ} in increasingly pebble dominated disks is shown in pebble fraction increments of 0.1 assuming Poisson distributed values for the uncertainty band (blue lines). The giant planet occurrence rate is obtained from \emph{in-situ} populations of $10^3$ systems per value of f\textsubscript{peb} after 2 Gyr. We consider planets to be giants here if the envelope mass exceeds the core mass. There is a general trend of increasing number of giants formed with larger pebble fractions. The frequency increases from 2-3\% up to about 11\% as the disks become more pebble dominated. This is in clear contrast to the results obtained with migration enabled shown before where, even in the pebbles-only case, the giant planet frequency is below 4\% (orange points). The envelope heating effect due to the accretion of planetesimals is easily overpowered when planets, unrealistically, form \emph{in-situ}. We thus identify the delay of runaway gas accretion \emph{combined} with strong inward migration to be responsible for the observed phenomenon of no giants forming in our nominal simulations of hybrid pebble-planetesimal disks.\\

		\subsection{Pebble isolation mass} \label{sec:pebble isolation mass}
		
		    \begin{figure}
    		\centering
            \includegraphics[width=\linewidth]{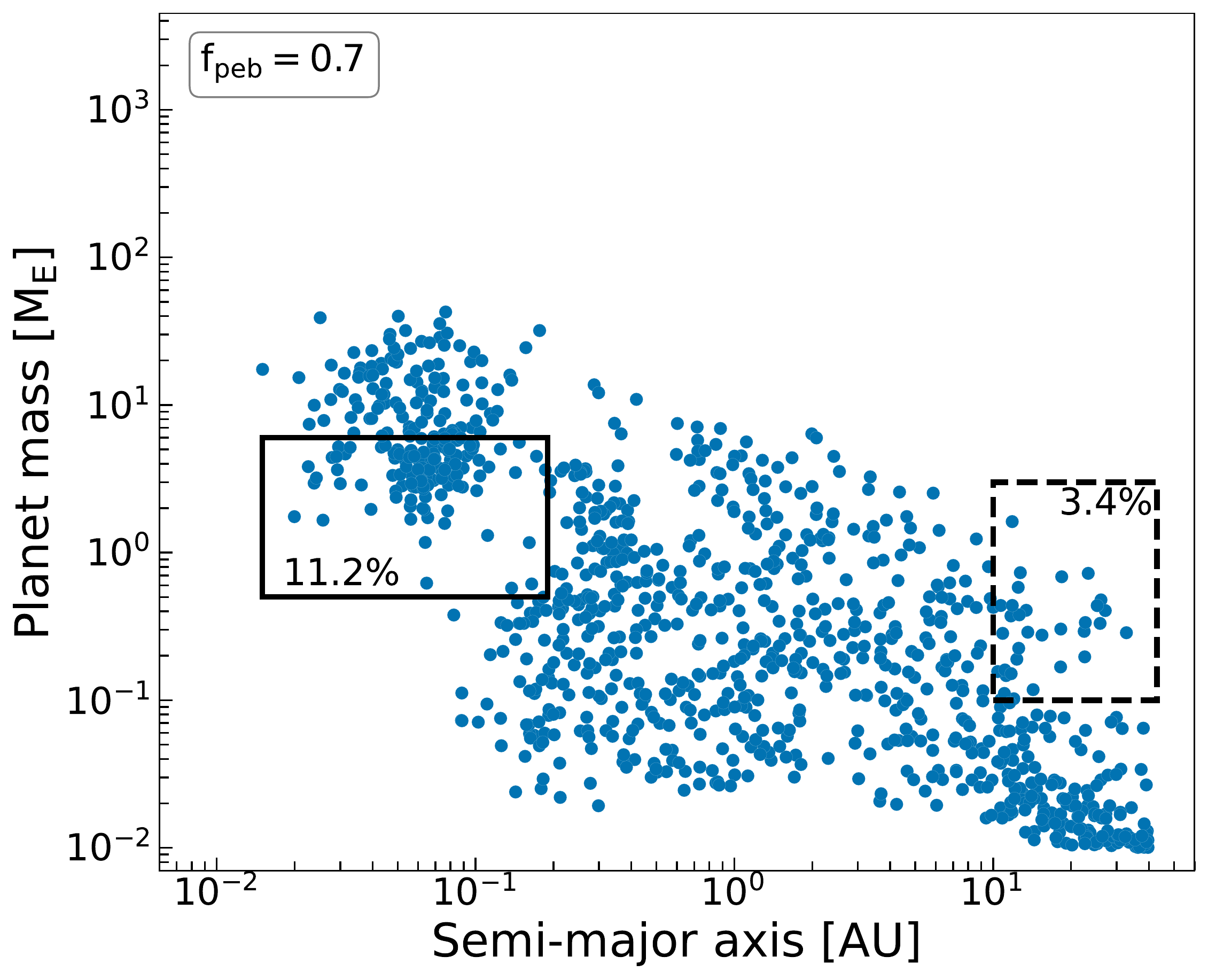}
                \caption{
                    Planet mass over semi-major axis diagram of the population for f\textsubscript{peb} = 0.7 with a pebble isolation mass that is double the value given in equation \eqref{eq:M_iso}. The solid-line box highlights planet masses of 0.5 to 6 M\textsubscript{E} in the inner disk region up to 0.2 AU. The dashed-line box highlights planet masses above 0.1 M\textsubscript{E} outside of 10 AU. The boxes are labelled with the percentage of planets in these regions.
                    }
                \label{fig:0p7_2miso}
            \end{figure}
		
		    Another possible influence to giant planet formation comes from the value of the pebble isolation mass. Since this mass sets an upper limit to pebble accretion, it could be too low for significant gas accretion to happen, especially in the inner regions of the disk. As already shown in Fig. \ref{fig:mvsa_ring}, giants can form in a pebbles-only setting in the outer disk, where M\textsubscript{iso} is large, and subsequently migrate closer in. Also in the \emph{in-situ} f\textsubscript{peb} = 0.7 case in Fig. \ref{fig:0p7_nomig}, the pebble isolation mass is only reached by planets inside of roughly 0.7 AU. Outside of that, pebble accretion is rather limited by the depletion of pebbles or the disk lifetime. In Fig. \ref{fig:0p7_2miso}, the population for f\textsubscript{peb} = 0.7 is shown with a doubled value of the pebble isolation mass. It is qualitatively indistinguishable from the nominal pebble-rich case in the inner disk and identical in the outer disk where planets do not reach pebble isolation anyway.  Even an overestimated value of M\textsubscript{iso} does not assist the formation of giants anywhere in a hybrid pebble-planetesimal disk. Close to the inner edge, where the pebble isolation mass is lowest and a change of M\textsubscript{iso} shifts planetary masses accordingly (see solid-line box), giant formation is unlikely due to inward migration. For this reason, giant formation models normally focus on initial orbital distances of several AU where we find the accretion heating of the envelope and inward migration to be the dominant mechanisms at play.

        \subsection{Pebble flux timing} \label{sec:pebble flux timing}

        \begin{figure}
    	\centering
            \includegraphics[width=\linewidth]{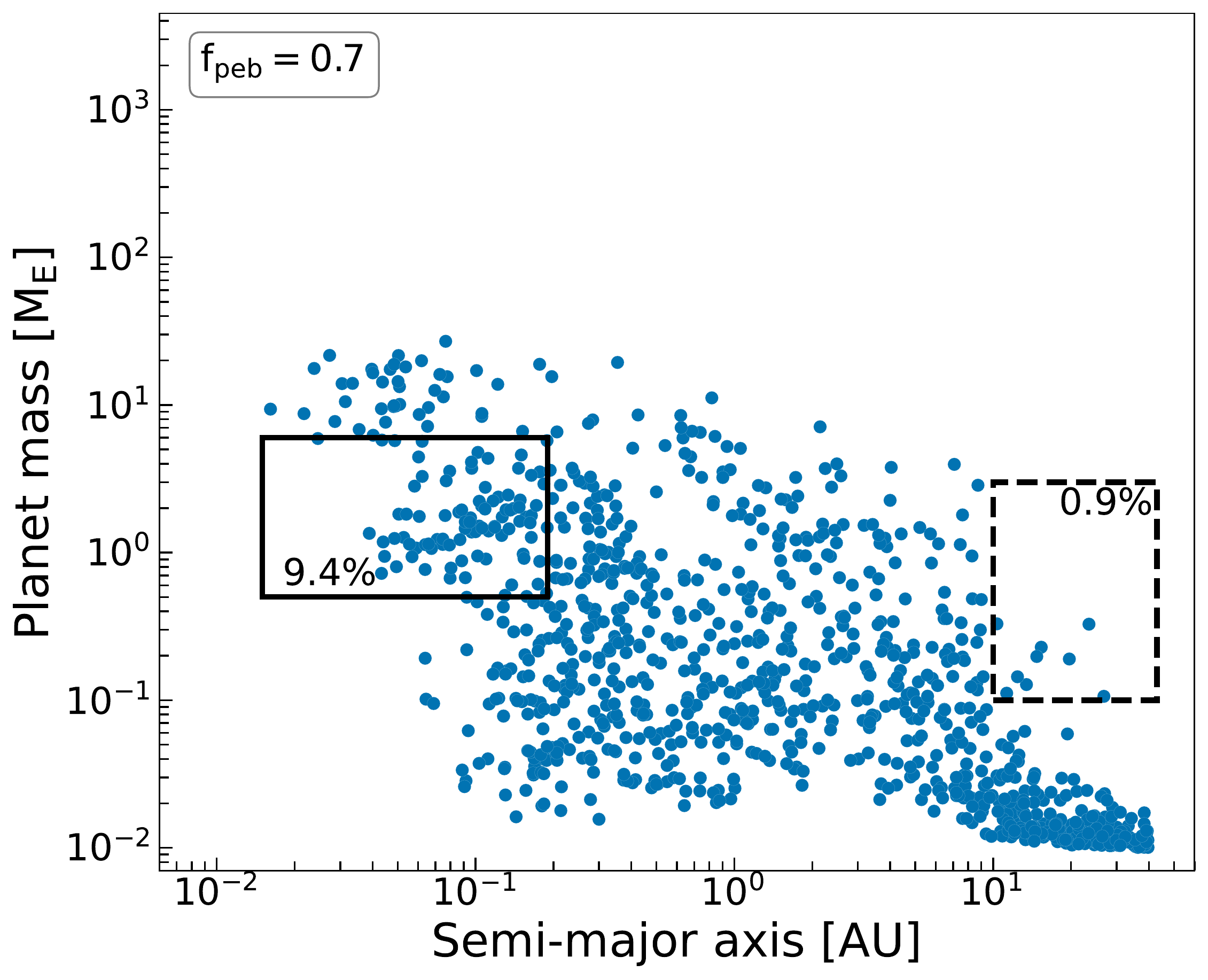}
            \caption{
                Planet mass over semi-major axis diagram of the population for f\textsubscript{peb} = 0.7 with a lowered pebble formation efficiency ($\epsilon_d=\epsilon_p=0.05$ instead of 0.5). The solid-line box highlights planet masses of 0.5 to 6 M\textsubscript{E} in the inner disk region up to 0.2 AU. The dashed-line box highlights planet masses above 0.1 M\textsubscript{E} outside of 10 AU. The boxes are labelled with the percentage of planets in these regions. 
                }
            \label{fig:0p7_lowepsilon}
        \end{figure}

        As mentioned in Sect. \ref{sec:pebble formation model}, the pebble growth radius sweeps through the disk within about 1 Myr in this model. There is however, observational evidence of pebbles in disks that are much older than that which suggests that a flux of pebbles could be present at later times. In \citet{levisonGrowingGasgiantPlanets2015}, it was shown that a lower pebble flux that is maintained for longer allows for the formation of giant planets. However, these findings were obtained from multi-planet simulations where dynamical interactions remove the smaller embryos, resolving the issue of forming many earth sized planets and no giants which was found in \citet{kretkeChallengesFormingSolar2014}. In our single-planet simulations, this exact interaction can not be replicated but the pebble flux timing is nevertheless relevant.\\
        We attempt to address the observation of pebbles at later times by arbitrarily reducing the pebble formation efficiency ($\epsilon_d=\epsilon_p=0.05$ instead of 0.5, see Sect. \ref{sec:pebble formation model}). This results in lower, later pebble fluxes because the pebble growth line moves slower due to the longer pebble growth timescales. The pebble growth radius now reaches the outer disk edge about a factor of 10 times later, extending the presence of pebbles to the order of the gas disk lifetime. In Fig. \ref{fig:0p7_lowepsilon}, the f\textsubscript{peb} = 0.7 simulation using the lower pebble formation efficiency is shown. Overall, we find a similar picture to the nominal simulation but with notably less planets that migrate to the inner disk regions (see solid-line box). This is explained by the later onset of pebble accretion and the lower magnitude of the pebble flux due to the lower gas surface densities towards the end of the disk lifetime. This leads to later and slower planet formation. As a consequence, the planet masses are now starting to be limited by the gas disk lifetime in the outer disk (see dashed-line box).\\
        Notably, the late pebble flux in this low efficiency scenario does not resolve the non-formation of giants in the hybrid simulations, even though migration is reduced.\\

        In this model, the lifetime of the pebble flux does not only depend on the pebble formation efficiency but also on the location of the outer disk edge. This is an intrinsic feature of pebble based planet formation. Since the disk size is a varied quantity in all the presented populations, the pebble flux lifetime is also varied. Within the probed parameter range, even the longest lived pebble fluxes evidently do not result in giant planets in the hybrid scenarios. Nevertheless, disk size is relevant for the formation of giant planets in the pebbles-only case as giant planets only form in disks of sufficient size corresponding to a characteristic radius of at least about 60 AU (see Fig. \ref{fig:rchar}). This is consistent with the lack of giants formed by pebbles in small disks in \citet{bruggerPebblesPlanetesimalsOutcomes2020}. Unsurprisingly, we also find a positive correlation of higher initial disk masses and the formation of giants (see Fig. \ref{fig:disks}).
        
	\section{Summary and conclusions} \label{sec:discussion and conclusion}
	
	    We combine a simple model of pebble formation and accretion with a global model of planet formation considering the accretion of planetesimals. Using a population synthesis approach for single planets, we investigate the effect of hybrid pebble-planetesimal disks on planet formation.\\
	    The main results obtained from populations of disks with different pebble fractions can be summarised as follows:
	    
	    \begin{itemize}
	        \item No giant planets are able to form in hybrid pebble-planetesimal disks, whereas planetesimals alone or pebbles alone form giants.
	        \item Inward migration is more prevalent when more pebbles are available because more planets grow to the point where they are subject to significant type-I migration.
	    \end{itemize}
	    
	    From the closer investigation of giant formation pathways we report the following findings:
	    
	    \begin{itemize}
	        \item Remaining planetesimal accretion after the pebble accretion phase adds sufficient energy to delay the onset of runaway gas accretion of massive cores in hybrid pebble-planetesimal environments.
	        \item Type-I migration acts strongly on giant planet candidates that do not immediately open a gap in the gas disk.
	        \item The combination of delayed runaway gas accretion and strong inward migration prevents the formation of giant planets in our simulations of hybrid pebble-planetesimal disks.
	    \end{itemize}
     
	    The simplicity of the pebble model and the use of single-embryo simulations allow us to disentangle the multitude of interdependent mechanisms acting in planet formation at the same time. On the other hand, this also prevents us from making final statements about the outcome of a more true-to-nature description of planet formation from dust all the way to multiple planets. Therefore, the above mentioned results do not imply that giant formation is generally impossible in this setting but they demonstrate the effects arising from the simultaneous accretion of pebbles and planetesimals and how they influence the formation pathway of planets fundamentally. The main conclusion we draw is that, in a combined pebble-planetesimal accretion scenario, planet formation is not necessarily boosted by the avenue of pebble accretion. Specifically for the formation of giant planets, we show that the accretion of pebbles as well as planetesimals can have a hindering effect and that the gap opening and the subsequent shift to the type-II migration regime is necessary for the survival of giant planets. This further underlines the importance of accretion heating for the correct calculation of gas accretion rates and the fact that orbital migration in general is a non-negligible process in planet formation. Note that this is also a consequence of the turbulent viscosity parameter $\alpha=0.002$ chosen in this work. In disks of lower viscosity, the transport of angular momentum in the disk is less efficient which leads to lower gas driven migration rates and lower gap opening masses. This means that the formation of giant planets might be suppressed less if $\alpha$ is low. Additionally, the prescriptions for orbital migration described in \ref{sec:orbital migration} do not include the thermal torque which could allow a higher fraction of planets to stay in the outer disk due to outward migration \citep{baumannInfluenceMigrationModels2020a,guileraImportanceThermalTorques2021}.\\
	    In their study of the formation of a planetary system considering pebble and planetesimal accretion, apart from not forming any giant planets, \citet{voelkelMultipleGenerationsPlanetary2022a} find a first generation of pebble-formed terrestrial planets which are accreted by the star due to efficient type-I migration. These hints at a possible detrimental effect of efficient pebble accretion on planet formation are complemented by our results.\\
        Regarding the proposed Jupiter formation scenario in \cite{alibertFormationJupiterHybrid2018}, our results confirm the plausibility of delayed runaway gas accretion in hybrid disks. However, the notion of a massive planetary core staying at the initial position for multiple Myr is clearly challenged by this work.\\
	   
	    In a more complete model, a number of additional effects are expected to influence the results found in this study. Since pebbles are relatively well coupled to the gas, the structure of the gas disk changes the pebble dynamics strongly. This is especially relevant when progressing from a single-planet scenario to the formation of multi-planetary systems. Planet-gas interactions are important here because massive planets can trap pebbles and effectively shield other growing planets from the pebble flux, leaving them in an accretion environment more akin to the planetesimals-only picture. In \citet{stammlerLeakyDustTraps2023} however, it was recently found that gaps in the disk might not be efficient traps for smaller pebbles and dust. This could still allow pebble accretion inside of massive outer planets. The accumulation of pebbles is also relevant in the context of the N-body interactions between the planets. For example, if a planet moves through a pile-up of pebbles caused by another planet, it can accrete a large amount of pebbles in a short time. The assumption of drift limited pebble formation and evolution is clearly no longer viable under these circumstances. Additionally, the gravitational interactions among multiple planets changes the migration behaviour, for instance due to mean motion resonances. As shown in Sect. \ref{sec:inward migration}, preventing the inward migration of the planet all the way to the inner disk edge can allow massive cores to form giants.\\
	    For these reasons, it is impossible to predict the outcome of (giant) planet formation in hybrid pebble-planetesimal disks in multi-planet population syntheses from these results but we expect the underlying mechanisms of delayed runaway gas accretion and increased orbital migration to persist.

    \begin{acknowledgements}
        We acknowledge the support from the Swiss National Science Foundation (SNSF) under grant 200020\textunderscore192038. We would like to thank the anonymous referee for the valuable comments and suggestions that helped us improve the manuscript.
    \end{acknowledgements}
 
	\bibliographystyle{bibtex/aa}
    \bibliography{Pebble_planetesimal_interplay.bbl}

    \begin{appendix} 
    \section{Initial disk properties}\label{app:initial disk properties}

        \begin{figure}
    	\centering
            \includegraphics[width=\linewidth]{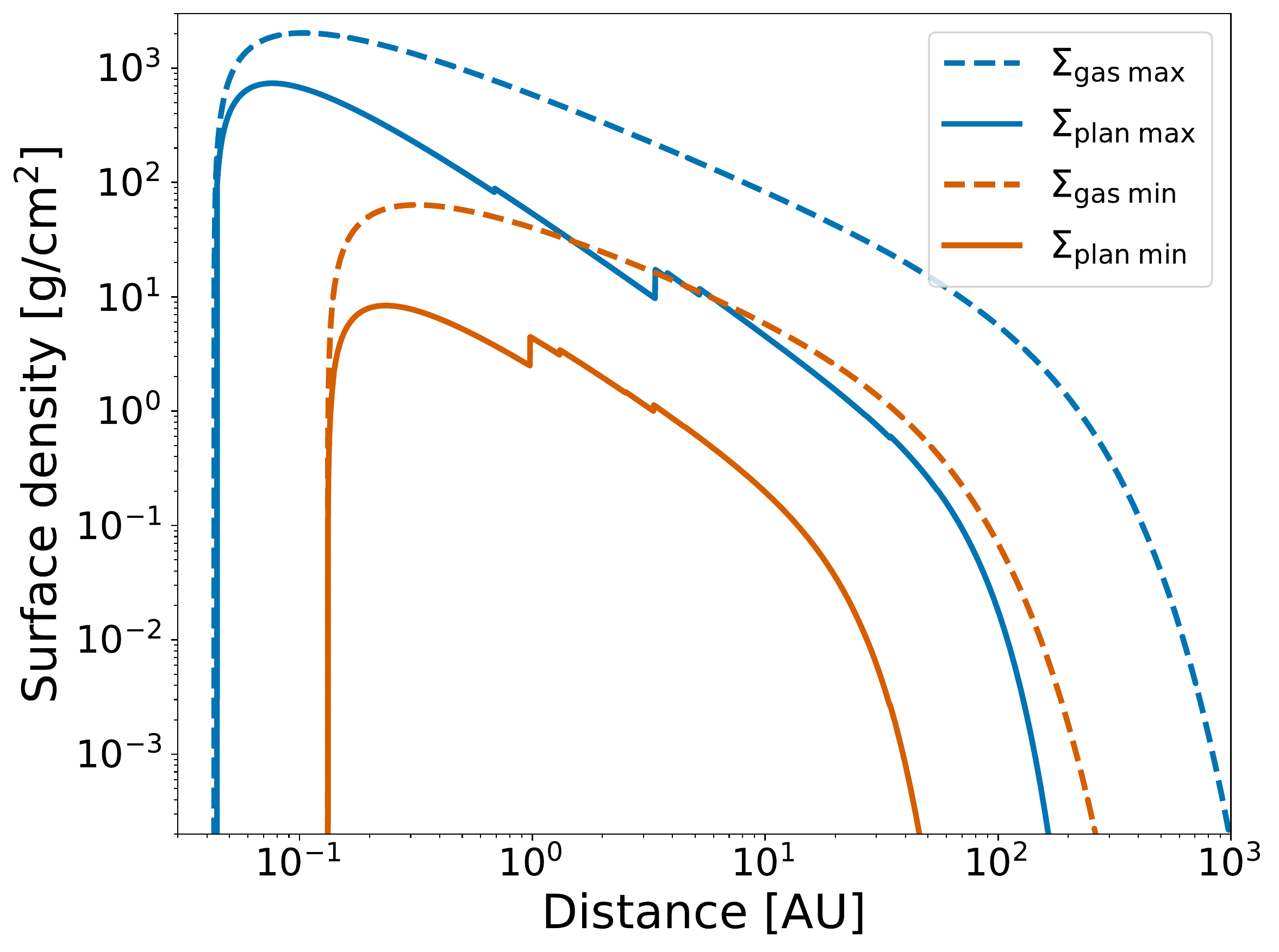}
                \caption{
                    Initial radial gas (dashed) and planetesimal (solid) surface density profiles. The blue (orange) lines correspond to the system with the most (least) massive planetesimal disk.
                    }
                \label{fig:profiles}
        \end{figure}
        
        \begin{figure}
    	\centering
            \includegraphics[width=\linewidth]{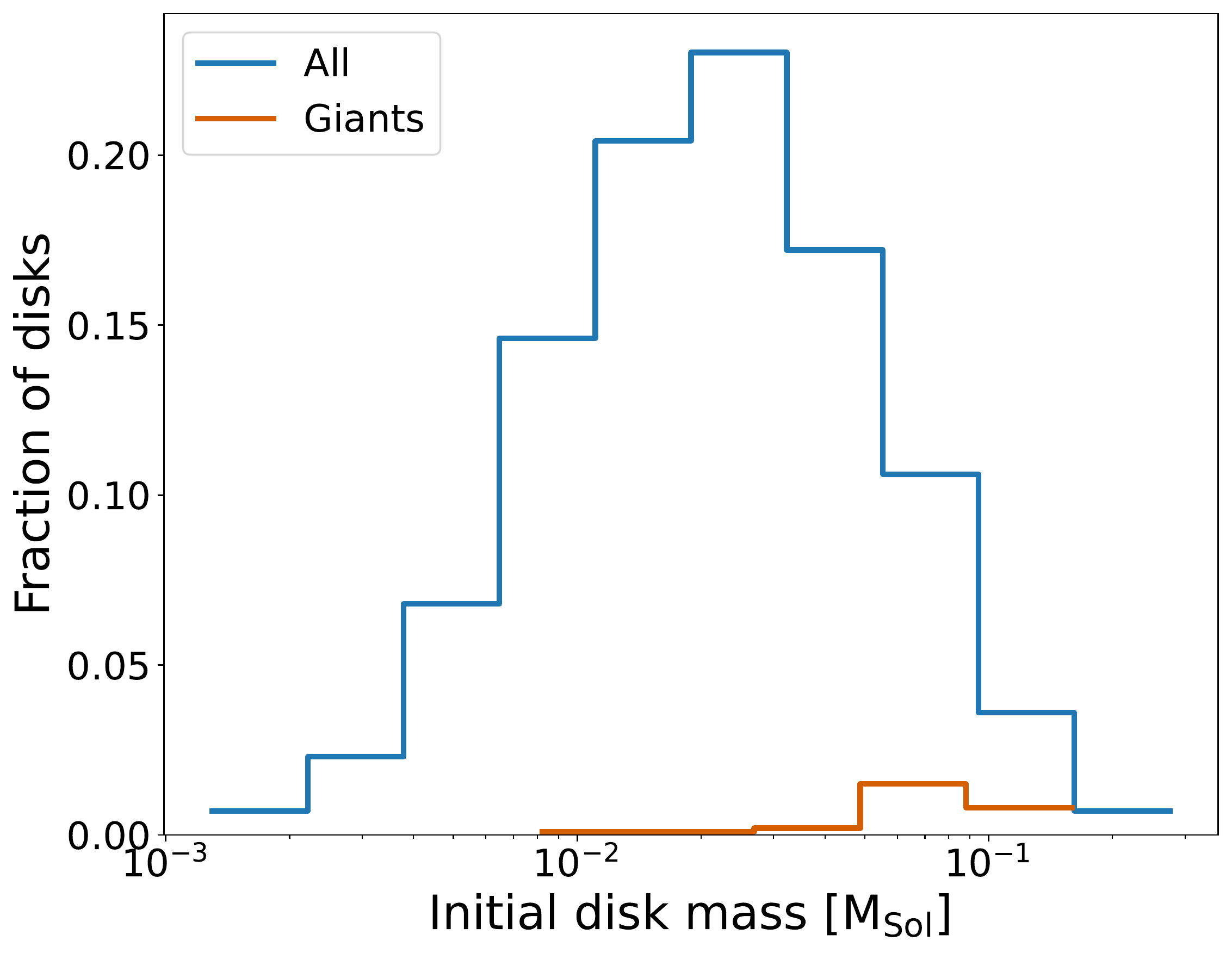}
                \caption{
                    Fraction of disks of a given initial gas disk mass. The full set of 1000 disks is shown in blue and the orange line shows the disks that form a giant planet in the f\textsubscript{peb} = 1 simulations.
                    }
                \label{fig:disks}
        \end{figure}

        \begin{figure}
    	\centering
            \includegraphics[width=\linewidth]{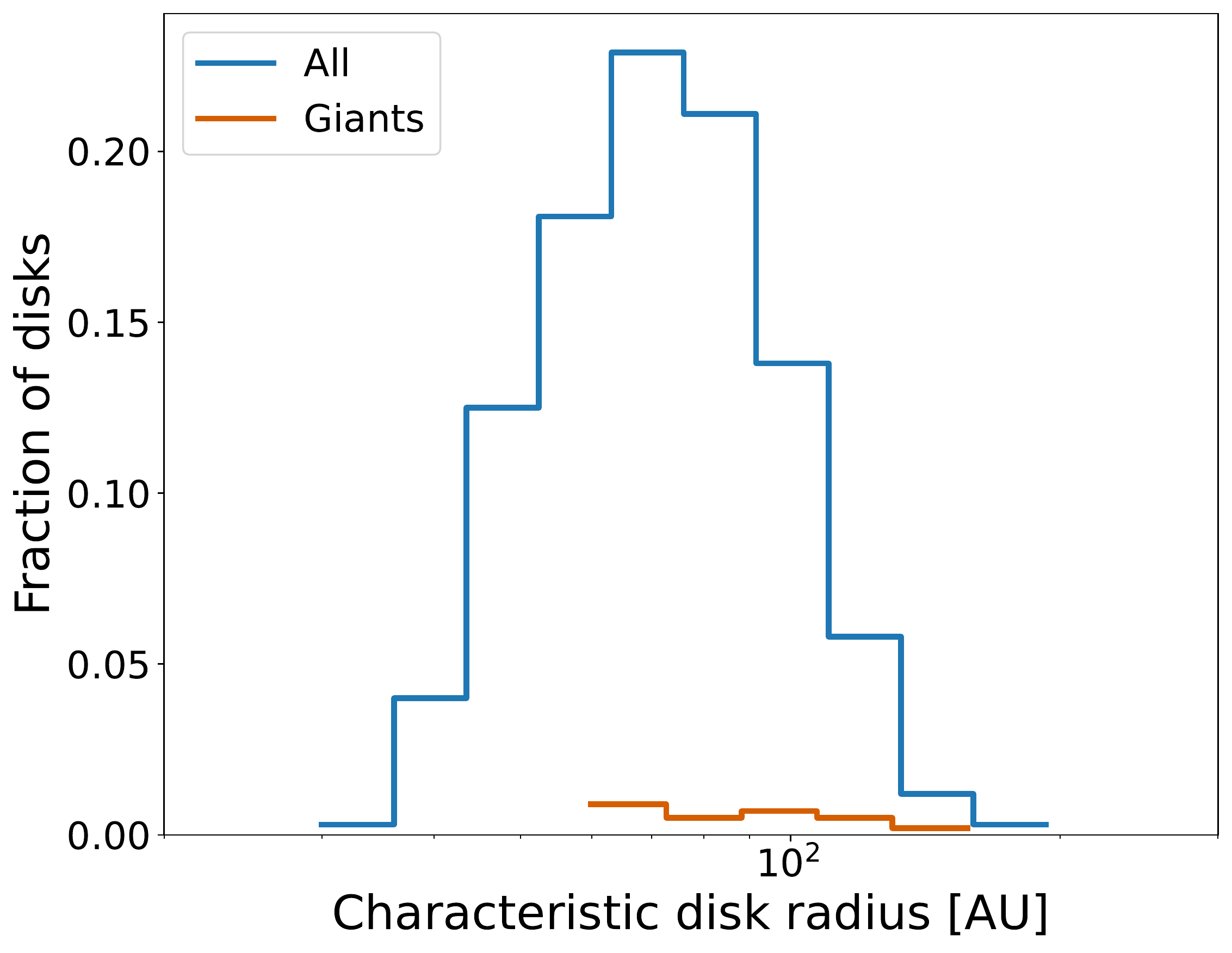}
                \caption{
                    Fraction of disks of a given initial characteristic radius.The full set of 1000 disks is shown in blue and the orange line shows the disks that form a giant planet in the f\textsubscript{peb} = 1 simulations.
                    }
                \label{fig:rchar}
        \end{figure}
        The initial radial gas and planetesimal surface density profiles are shown in Fig. \ref{fig:profiles}. We show the most (blue) and least (orange) massive \emph{planetesimal} disks. The planetesimal disk mass is a function of the gas disk mass, the size of the gas disk, and the solids-to-gas ratio. Hence the disks shown here are not necessarily also the most or least massive gas disks.\\
        The distribution of initial gas disk masses is shown in Fig. \ref{fig:disks} (blue). Note that the total number of disks considered in this work is 1000 and that the stellar mass is fixed to one solar mass. We find a positive correlation of high initial disk masses and the formation of giants in the pebbles-only scenario (orange).\\
        As described in Sect. \ref{sec:population synthesis outcomes}, the characteristic gas disk radius is a derived quantity. The resulting distribution of characteristic gas disk sizes is shown in Fig. \ref{fig:rchar} (blue). We find no clear correlation of initial disk sizes, and the associated longer pebble flux lifetimes, with the formation of giant planets in the f\textsubscript{peb} = 1 simulations (orange). However, giant planets only form in disks of sufficient size corresponding to a characteristic radius of at least about 60 AU.

    \end{appendix}
\end{document}